\documentclass[11pt,a4paper]{scrartcl}
\usepackage[a4paper, includefoot, total={159mm,242mm}]{geometry}
\usepackage[T1]{fontenc} 
\usepackage[utf8]{inputenc} 
\usepackage[english]{babel} 
\usepackage{lmodern} 
\usepackage{amsmath,amsfonts,amssymb} 
\usepackage{microtype} 
\usepackage{graphicx} 
\usepackage{caption} 
\usepackage{afterpage} 
\allowdisplaybreaks 

\DeclareOldFontCommand{\rm}{\normalfont\rmfamily}{\mathrm}
\DeclareOldFontCommand{\sf}{\normalfont\sffamily}{\mathsf}
\DeclareOldFontCommand{\tt}{\normalfont\ttfamily}{\mathtt}
\DeclareOldFontCommand{\bf}{\normalfont\bfseries}{\mathbf}
\DeclareOldFontCommand{\it}{\normalfont\itshape}{\mathit}
\DeclareOldFontCommand{\sl}{\normalfont\slshape}{\@nomath\sl}
\DeclareOldFontCommand{\sc}{\normalfont\scshape}{\@nomath\sc}

\begin{document}

\renewcommand{\theequation}{\thesection.\arabic{equation}}

\newcommand{\abcdon}{\addtocounter{equation}{1}%
 \setcounter{help}{\value{equation}}%
 \setcounter{equation}{0}%
 \renewcommand{\theequation}{\thesection.\arabic{help}\alph{equation}}}

\newcommand{\appendixon}
  {\setcounter{section}{0}
   \renewcommand{\thesection}{\Alph{section}}}
\newcommand{\appendixoff}
  {\renewcommand{\thesection}{\arabic{section}}}

\newcommand{\namedappendix}[1]
 {\setcounter{equation}{0}
  \renewcommand{\thesection}{\Alph{section}}
  \refstepcounter{section}
  \section*{Appendix~\Alph{section}: #1}
   \setcounter{table}{0}
  \addcontentsline{toc}{chapter}{\protect\numberline{\thesection}#1}
  \vspace{1.0ex} \par}

\newcommand{\eco}{\tilde{E}}
\newcommand{\eln}{ \mathcal{l}n}
\newcommand{\no}{\nonumber \\ }
\newcommand{\gqq}{\gamma^{(0)}_{S,qq}}
\newcommand{\gqg}{\gamma^{(0)}_{S,qg}}
\newcommand{\ggq}{\gamma^{(0)}_{S,gq}}
\newcommand{\GGG}{\gamma^{(0)}_{S,gg}}
\newcommand{\Gqq}{\gamma^{(1)}_{S,qq}}
\newcommand{\Gqg}{\gamma^{(1)}_{S,qg}}
\newcommand{\Ggq}{\gamma^{(1)}_{S,gq}}
\newcommand{\Ggg}{\gamma^{(1)}_{S,gg}}
\newcommand{\aqq}{a_{S,qq}^{(1)}}
\newcommand{\aqg}{a_{S,qg}^{(1)}}
\newcommand{\agq}{a_{S,gq}^{(1)}}
\newcommand{\agg}{a_{S,gg}^{(1)}}
\newcommand{\Aqq}{a_{S,qq}^{(2)}}
\newcommand{\Aqg}{a_{S,qg}^{(2)}}
\newcommand{\Agq}{a_{S,gq}^{(2)}}
\newcommand{\Agg}{a_{S,gg}^{(2)}}
\newcommand{\bqq}{b_{S,qq}^{(1)}}
\newcommand{\bqg}{b_{S,qg}^{(1)}}
\newcommand{\bgq}{b_{S,gq}^{(1)}}
\newcommand{\bgg}{b_{S,gg}^{(1)}}
\newcommand{\gs}{\left( \frac{g^2}{16 \pi^2} \right)}
\newcommand{\gsgs}{\left( \frac{g^2}{16 \pi^2} \right)^2}
\newcommand{\apsqq}{a_{PS,qq}^{(2)}}
\newcommand{\ansqq}{a_{NS,qq}^{(1)}}
\newcommand{\bnsqq}{b_{NS,qq}^{(1)}}
\newcommand{\gnsqq}{\gamma_{S,qq}^{(0)}}
\newcommand{\gpsqq}{\gamma_{PS,qq}^{(1)}}
\newcommand{\lnp}{ \ln \left(- \frac{p^2}{\mu^2} \right) }
\newcommand{\lnpp}{ \ln^2 \left(- \frac{p^2}{\mu^2} \right) }
\newcommand{\ps}{p\hspace{-0.42em}/\hspace{0.1em}}
\newcommand{\qs}{q\hspace{-0.42em}/\hspace{0.1em}}
\newcommand{\Ds}{\Delta\hspace{-0.52em}/\hspace{0.1em}}
\newcommand{\as}{a\hspace{-0.5em}/\hspace{0.1em}}
\newcommand{\bs}{b\hspace{-0.5em}/\hspace{0.1em}}
\newcommand{\cs}{c\hspace{-0.5em}/\hspace{0.1em}}
\newcommand{\ds}{d\hspace{-0.5em}/\hspace{0.1em}}
\newcommand{\ks}{k\hspace{-0.52em}/\hspace{0.1em}}

\newcommand{\e}{\varepsilon}
\newcommand{\fe}{\frac{1}{\varepsilon}}
\newcommand{\fee}{\frac{1}{\varepsilon^2}}
\newcommand{\dx}{\delta(1-x)}
\newcommand{\lmx}{\ln (1-x)}
\newcommand{\lx}{\ln x}
\newcommand{\li}{{\rm Li}_2}

\newcommand{\pld}{\left( \frac{1}{1-x} \right)_+}
\newcommand{\lpld}{\left( \frac{\ln (1-x)}{1-x} \right)_+}
\newcommand{\llpld}{\left( \frac{\ln^2 (1-x)}{1-x} \right)_+}
\newcommand{\MSb}{${\rm \overline{MS}}$}
\newcommand{\MSbs}{${\rm \overline{MS}}$ scheme}

\newcommand{\lc}{\left\{}
\newcommand{\rc}{\right\}}

\newcommand{\sumi}{\sum_{i=0}^{m-3}}
\newcommand{\sumj}{\sum_{j=0}^{m-3}}
\newcommand{\ddr}{(\Delta r)}
\newcommand{\dds}{(\Delta s)}
\newcommand{\ddp}{(\Delta p)}
\newcommand{\ddq}{(\Delta q)}

\hyphenation{dia-grams}
\setcounter{page}{0}

\thispagestyle{empty}

\hfill INLO--PUB--6/95

\hfill NIKHEF-H/95-031

\hfill November 1995 (revised)

\vspace*{1cm}
\begin{center}
\large{\bf
The Calculation of the
Two-Loop Spin Splitting Functions $P_{ij}^{(1)}(x)$
}
\end{center}

\vspace*{2cm}
\begin{center}
R.~Mertig\footnote{rolfm@gluonvision.com}  \\~\\~
NIKHEF-H, \\ P.O. Box 41882 \\  1009 DB Amsterdam \\ The Netherlands
\end{center}

\vspace*{.5cm}
\begin{center}
and
\end{center}
\vspace*{.5cm}

\begin{center}
W.L.\ van Neerven \\~\\~
Instituut--Lorentz \\ University of Leiden \\ P.O. Box 9506 \\
2300 RA Leiden \\ The Netherlands
\end{center}

\vspace*{2cm}

\noindent
{\bf Abstract}

\vspace*{8mm} \noindent
We present the calculation of the two-loop spin splitting
functions $P_{ij}^{(1)}(x)\; (i,j = q,g)$ contributing to the
next-to-leading order corrected spin structure function
$g_1(x,Q^2)$. These splitting functions,
which are presented in the \MSbs,
 are derived from the order
$\alpha_s^2$ contribution to the anomalous dimensions
$\gamma_{ij}^{m} \; (i,j = q,g)$.
The latter correspond to the local operators
which appear
in the operator product expansion of
two electromagnetic currents.
Some of the properties of the anomalous dimensions will be
discussed. In particular our findings are in agreement with 
the supersymmetric relation
$\gamma_{qq}^{m} + \gamma_{gq}^{m} -
\gamma_{qg}^{m} - \gamma_{gg}^{m} = 0$
up to order $\alpha_s^2$.

\newpage
\section{Introduction}

During the last few years there has been a great deal of
activity in the area of polarized lepton-hadron physics
both from the experimental as well as the theoretical side.
This interest started with the discovery of the EMC-experiment
\cite{c1} that the Ellis-Jaffe sum rule \cite{c2},
which represents the first moment of the spin structure
function $g_1(x,Q^2)$, was violated by the combined
SLAC-EMC data \cite{c1,c3}. This discrepancy between
theory and experiment, also called the ``spin crisis'',
came as a great surprise because one expected that sum rules
derived in the context of the constituent quark model,
which is valid at low
energy scales, should also hold at large energy
scales characteristic of the current quark (parton) regime.
In particular the constituent quark model assumes that the
spin of the proton can be mainly attributed to its valence
quarks and the sea quark contribution is negligible small.
This assumption leads to a value of the Ellis-Jaffe sum rule which
is appreciably larger than the one found by experiment.
Although more recent experiments \cite{c4,c5,c6} lead to a result which
is closer to the theoretical prediction the discrepancy is still large
enough to warrant explanation.

Many theorists have tried to explain the above discrepancy
(for recent reviews see \cite{c7}) in the framework of perturbative
and also non-perturbative QCD. From this theoretical work one can draw the
conclusion that the interpretation of the spin structure function
$g_1(x,Q^2)$, using the ideas of the operator product
expansion (OPE) and
the QCD improved parton model, is not as simple as that
given to the structure functions which show up in unpolarized
lepton-hadron scattering.
In particular singlet the axial vector operator is renormalized due to the
Adler-Bell-Jackiw anomaly.
Therefore the interpretation that the
polarized parton densities represent the spin carried by the
corresponding partons does not hold anymore.
Fortunately this operator cancels in the Bjorken sum rule \cite{c8}
so that the latter has a more reliable theoretical basis.
It is therefore no surprise that its result is in agreement with
recent data \cite{c4,c5,c6}.
The above theoretical work also led to many different parametrizations
of the parton densities in terms of which the spin structure function
$g_1(x,Q^2)$ can be expressed. One of the key issues is the role
of the gluon density which can account for the negative contribution
to the Ellis-Jaffe sum rule depending on the chosen scheme.
However, if one wants to give a complete next-to-leading order
(NLO) description of $g_1(x,Q^2)$, and not only its first moment,
one needs a full knowledge of the order $\alpha_s$ coefficient
functions, which are known
(see e.~g.~\cite{X9,X10,X11})
 and the order
$\alpha_s^2$ corrected Altarelli-Parisi (AP) spin splitting
functions $ P_{ij} \;  (i,j = q,g)$.
The lowest order AP-splitting functions $ P_{ij}^{(0)}$ have been
calculated in \cite{X12} and \cite{X13} respectively using different methods.
In \cite{X12} the operator product expansion (OPE)
techniques are applied to obtain the anomalous dimensions
of the composite
operators appearing in the spin dependent part of the
current-current correlation function.
The latter appears in the
expression for the deep inelastic cross section.
The authors in \cite{X13} have
used the parton model approach. The NLO (order $\alpha_s^2$) splitting
functions $ P_{qq}^{(1),S}$ and $ P_{qg}^{(1)}$ have been
computed in \cite{X9} using the standard techniques of perturbative QCD.
They emerge while performing mass factorization of the order
$\alpha_s^2$ corrected parton cross sections of the processes
$\gamma^* q$ and $\gamma^* g$ which contribute to the
deep inelastic spin structure function.
Unfortunately the remaining splitting functions
$ P_{gq}^{(1)}$ and $ P_{gg}^{(1)}$
could not be
obtained in this way since they do not show up in the
mass factorization of order $\alpha_s^2$ corrected parton cross
sections. This can be traced back to the phenomenon that there is no
direct coupling of the virtual photon $\gamma^*$ or any other
electroweak vector boson to the gluon. Therefore
$ P_{gq}^{(1)}$ and $ P_{gg}^{(1)}$
will appear in the mass factorizaton of the order
$\alpha_s^3$ corrected parton cross sections which are very
difficult to calculate. In order to avoid the above complication
we will resort to the standard OPE techniques to
calculate the missing splitting functions which are derived
from the inverse Mellin transform of the anomalous dimensions of
composite operators.
Finally we want to emphasize that in this paper we
will limit ourselves to the presentation and the discussion
of the properties of the spin anomalous dimensions (AP
splitting functions) only. The effect of their contribution to
the analysis of the spin structure function $g_1(x, Q^2)$
lies beyond the scope of our paper. Notice that this
analysis is the same as the one performed for the spin
averaged structure function $F_2(x, Q^2)$ since the leading
contribution to both structure functions originate from
twist two operators only (for a discussion see section
2 below). Therefore the results for $g_1(x, Q^2)$ can be obtained
from the existing programmes for $F_2(x, Q^2)$. In
the latter one has to replace the spin averaged anomalous
dimensions (AP splitting functions) and the coefficient
functions by their spin analogues presented in this
paper.

The paper is organized as follows.
In section 2 we introduce our notations and present a short discussion
of the composite twist-2 operators contributing to the
spin structure function $g_1(x,Q^2)$.
Here we also derive the general form of the renormalized
and unrenormalized operator matrix elements (OMEs)
where the operators are
sandwiched between polarized quark and gluon states.
The calculation of the OMEs is presented in section 3,
from which one extracts the anomalous
dimensions and the AP splitting functions which are
presented in the \MSbs.
Further we give the lowest order coefficient functions of
$g_1(x,Q^2)$ in the same scheme.
The properties of the anomalous dimensions are discussed in section 4.
In Appendix A one can find the operator vertices needed
for the computation of the operator matrix elements in section 3.
The tensorial reduction of the Feynman integrals which show
up in the calculation is discussed in Appendix B.

\section{Operators contributing to
the spin structure function \newline $g_1(x,Q^2)$
}

In this section we specify the composite
operators which appear in the light-cone expansion of two
electromagnetic currents.
Furthermore we present the operator matrix elements
(OMEs) as a power series in the strong coupling constant.
The coefficients of the perturbation series are determined
by the renormalization group (Callan-Symanzik) equations.
We will write the OMEs in the most general way so that they
can be used to extract the anomalous dimensions of the
composite operators.
The light-cone expansion of two electromagnetic currents is
given in \cite{X12} and reads as follows
\begin{eqnarray}
J_\mu(z) J_\nu(0)
&
\stackrel{z^2 \rightarrow 0}{ \simeq}
&
(-g_{\mu \nu} \Box + \partial_\mu \partial_\nu)
\frac{1}{z^2-i \varepsilon z_0}
\sum_{m=0}^\infty \sum_i
C_{i,1}^{m}(z^2 - i \varepsilon z_0,\mu^2,g)
\nonumber \\
&&
z_{\mu_1} \cdots z_{\mu_m} O_i^{\mu_1 \cdots \mu_m}(0)
- \left(g_{\mu \mu_1} g_{\nu \mu_2} \Box -
g_{\mu \mu_1} \partial_\nu \partial_{\mu_2} -
g_{\nu \mu_2} \partial_\mu \partial_{\mu_1}
\right.
\nonumber \\
&&
\left.
+ g_{\mu \nu} \partial_{\mu_1} \partial_{\mu_2}
\right)
\sum_{m=2}^\infty
\sum_i
C_{i,2}^{m} (z^2 - i \varepsilon z_0, \mu^2, g)
z_{\mu_3} \cdots z_{\mu_m} O_i^{\mu_1  \cdots \mu_m}(0)
\nonumber \\
&&
- i \epsilon_{\mu \nu \lambda \mu_1} \partial^{\lambda}
\frac{1}{z^2-i \varepsilon x_0}
\sum_{m=1}^\infty \sum_i E_{i,1}^{m}(z^2
- i \varepsilon z_0,\mu^2,g)
\nonumber \\
&&
\phantom{
- i \epsilon_{\mu \nu \lambda \mu_1} \partial^{\lambda}
\frac{1}{z^2-i \varepsilon z_0}
\sum_{m=1}^\infty \sum_i
}
\times
z_{\mu_2} \cdots z_{\mu_m} R_i^{\mu_2  \cdots \mu_m}(0).
\label{2.1}
\end{eqnarray}
In the above we only consider the contribution of twist-2 operators.
The index $i$ of the locally gauge invariant operators
$O_i^{\mu_1 \cdots \mu_m}$ and
$R_i^{\mu_2  \cdots \mu_m}$
stands for the representation of the flavour
group $SU(n_f)$. Notice that the operators are also irreducible
representations of the Lorentz group which means that they are traceless
and symmetric in the Lorentz indices ${\mu_1 \cdots \mu_m}$.
The Wilson coefficient functions, denoted by $C_{i,k}^{m} \; (k=1,2)$
and $E_{i,1}^{m}$, can be expressed into a perturbation
series in the gauge (strong) coupling constant $g$.
Notice that all the above quantities are renormalized
which is indicated by the
renormalization scale $\mu$.
The product of the two electromagnetic currents appear in the
hadronic tensor defined in polarized deep inelastic lepton-hadron
scattering which is given by
\begin{eqnarray}
W_{\mu \nu}(p,q,s) &=& \frac{1}{4 \pi} \int d^4 z\; e^{i q z}
\langle p,s | J_\mu(z) J_\nu(0) | p,s \rangle
\nonumber \\
&=& W_{\mu \nu}^S(p,q) + i  W_{\mu \nu}^A(p,q,s).
\label{2.2}
\end{eqnarray}
Here $p$ and $s$ denote the momentum and spin of the hadron
respectively and $q$ stands for the virtual photon momentum.
The symmetric part of the hadronic tensor is given by
\begin{eqnarray}
W_{\mu \nu}^S(p,q) &=&
\left (-g_{\mu \nu} + \frac{q_\mu q_\nu}{q^2} \right  )
 F_1(x, Q^2) + \left (p_\mu - \frac{p \, q}{q^2} q_\mu \right )
\left (p_\nu - \frac{p \, q}{q^2} q_\nu \right ) \frac{F_2(x,Q^2)}{p \, q},
\label{2.3}
\end{eqnarray}
while the antisymmetric part is equal to
\begin{eqnarray}
W_{\mu \nu}^A(p,q,s)
&=& - \frac{m}{p \, q} \,
\epsilon_{\mu \nu \alpha \beta} q^\alpha \,
\left[ s^\beta g_1(x,Q^2) +
(s^\beta - \frac{s \, q}{p \, q} p^\beta)  \, g_2(x,Q^2)\right],
\label{2.4}
\end{eqnarray}
with the properties $s \cdot p = 0$, $s^2 = -1$ and
$m$ denotes the mass of the hadron.
The Bjorken scaling variable is given by $x = Q^2 / (2 p \, q)$ and
$Q^2 = - q^2 > 0$. The spin averaged structure functions
are denoted by $F_k(x,Q^2) (k=1,2)$.
In polarized electroproduction one has in addition
the longitudinal spin structure function $g_1(x,Q^2)$
 and the transverse  spin structure function $g_2(x,Q^2)$.
The twist-2 operators $O_i^{\mu_1 \cdots \mu_m}(0)$ corresponding
to the spin averaged structure functions are given in the literature
and their anomalous dimensions have been calculated up to
two-loop order
\cite{X14}--\cite{X17}.
The twist-2 operators contributing to the spin structure
functions are given by
\cite{X12}
\begin{eqnarray}
R_{NS,q}^{\mu_1 \cdots \mu_m}(z)
&= &
i^m \;  S  \;
 \left\{(\bar{\psi}(z) \gamma_5 \gamma^{\mu_1}
D^{\mu_2} \cdots D^{\mu_m} \frac{1}{2} \lambda_i \psi (z)) -
({\rm traces})
\right\},
\label{2.5}
\\
R_{S,q}^{\mu_1 \cdots \mu_m}(z)
&=&
i^m \;  S  \;
 \left\{(\bar{\psi}(z) \gamma_5 \gamma^{\mu_1}
D^{\mu_2} \cdots D^{\mu_m} \psi (z)) -
({\rm traces})
\right\},
\label{2.6}
\\
R_{S,g}^{\mu_1 \cdots \mu_m}(z)
&=&
 i^m \;  S  \;
 \left\{
\frac{1}{2} \epsilon^{\mu_1 \alpha \beta \gamma}
Tr(F_{\beta \gamma}(z) D^{\mu_2} \cdots D^{\mu_{m-1}}
   F_{\alpha}^{\mu_m}(z)
  ) - ({\rm traces})
\right\}.
\label{2.7}
\end{eqnarray}
The symbol $S$ in front of the curly brackets
stands for the symmetrization of the indices
$\mu_1 \cdots \mu_m$ and $\lambda_i$ is the flavour group
generator of $SU(n_f)$. The quark and the gluon field tensor
are given by $\psi(z)$ and $F_{\mu \nu}^a(z)$ respectively
and $F_{\mu \nu} = F_{\mu \nu}^a T^a$ where $T^a$
stands for the generator of the colour group $SU(N)$ $(N=3)$.
The covariant derivative is given by
$D_\mu = \partial_\mu + i g T^a A^a_\mu(x)$ where
$A_\mu^a(z)$ denotes the gluon field.
{}From eqs.~(\ref{2.5})--(\ref{2.7}) one infers that with respect
to the flavour group one can distinguish the local operators in a
non-singlet part represented by $R_{NS,q}$ and in a singlet
part consisting of $R_{S,q}$ and $R_{S,g}$.

In the Bjorken limit
($Q^2 \rightarrow \infty, x = Q^2/(2 p\, q)$  fixed) the
current-current correlation function in (\ref{2.2}) is dominated
by the light cone $z^2=0$ so that it is justified to make a light cone
expansion for the product of the two electromagnetic currents.
When $Q^2 \rightarrow \infty$ the leading contribution of
$g_1(x, Q^2)$ consists of the twist-2 operators listed in
(\ref{2.5})--(\ref{2.7}) whereas $g_2(x, Q^2)$ also
receives contributions of twist-3 operators which are not given in the
expansion in eq.~(\ref{2.1}). Since we are only interested
in the longitudinal spin structure function
$g_1(x,Q^2)$ we can limit ourselves to the renormalization of
the twist-2 operators mentioned in (\ref{2.5})--(\ref{2.7}).
Inserting the light cone expansion for $J_\mu (z) J_\nu(0)$ in
(\ref{2.2}) one can derive the following relation
\begin{equation}
\int_0^1 d x \;  x^{m-1} g_1(x, Q^2) =
\sum_i A_i^{m}(p^2, \mu^2,g)
\tilde{E}_{i,1}^{m}(Q^2,\mu^2,g)\;; \qquad
m\;\;{\rm odd}.
\end{equation}
The left-hand side of the above equation stands for the Mellin
transform of $g_1(x,Q^2)$ and the right-hand side is given by the
operator matrix element (OME).
\begin{equation}
\langle p,s | R_i^{\mu_1 \cdots \mu_m} (0) | p,s \rangle
=
i^m A_i^{m}(p^2,\mu^2,g) S \left\{
(s^{\mu_1} p^{\mu_2} \cdots p^{\mu_m}) - ({\rm traces}) \right\},
\end{equation}
with $i = $NS,S, and $\tilde{E}_i^{m}$ stands for the
coefficient function.
\begin{eqnarray}
\tilde{E}_i^{m}(Q^2,\mu^2,g) &=&
-\frac{1}{4} (Q^2)^{m}
\nonumber \\
&&
\times
\left(\frac{\partial}{\partial q^2}\right)^{m-1}
\int d^4 z e^{i q z}
\frac{1}{z^2 - i \varepsilon z_0} E_{i,1}^{m}(z^2-
i \varepsilon z_0, \mu^2,g).
\label{2.10}
\end{eqnarray}
The $Q^2$-evolution of the spin structure function is determined
by the anomalous dimensions of the composite operators in
eqs.~(\ref{2.5})--(\ref{2.7}). They are obtained from the
renormalized partonic OMEs
\begin{equation}
\langle j,p,s|R_{k,i}^{\mu_1 \cdots \mu_m}|j,p,s \rangle
=
A_{k,ij}^{m}(p^2,\mu^2,g) S \left\{(s^{\mu_1} p^{\mu_2} \cdots
p^{\mu_{m_1}}) - ({\rm traces}) \right\},
\label{2.11}
\end{equation}
where now the quark and gluon operators are sandwiched between
quark and gluon states.
The indices in (\ref{2.11}) stand for $k = NS,S$ and
$i=q,g$; $j = q,g$.
The $A_{k,ij}^{m}$ are derived from
the Fourier transform into momentum space of the connected
Green's functions
\begin{equation}
\langle 0|T(\bar{\phi_j}(x)
R_{k,i}^{\mu_1 \cdots \mu_m}(0) \phi_j(y)|0 \rangle_c,
\label{2.12}
\end{equation}
where the external lines are amputated.
The fields $\phi_i(x)$ stand either for the quark fields
$\psi(x)$ or for the gluon fields $A_\mu^a(x)$.
The renormalized partonic OMEs
satisfy the Callan-Symanzik equations
\begin{equation}
\left[ \mu \frac{\partial}{\partial \mu} +
\beta(g) \frac{\partial}{\partial g} +
\delta(\alpha,g) \frac{\partial}{\partial \alpha} +
\gamma_{NS,qq}^{m}(g)\right] A_{NS,qq}^{(m)} (p^2,\mu^2,g,\alpha) = 0,
\label{2.13}
\end{equation}
and
\begin{equation}
\left[ \left (\mu \frac{\partial}{\partial \mu} +
\beta(g) \frac{\partial}{\partial g}
+ \delta(\alpha,g) \frac{\partial}{\partial \alpha}
\right ) \delta_{ij}
+
\gamma_{S,ij}^{m}(g)\right] A_{S,jk}^{(m)} (p^2,\mu^2,g,\alpha) = 0.
\label{2.14}
\end{equation}
Here $\beta(g)$ denotes the $\beta$-function which
in QCD is given by the following series expansion
\begin{eqnarray}
\beta(g) &=& -\beta_0 \frac{g^3}{16 \pi^2}
- \beta_1 \frac{g^5}{(16 \pi^2)^2} + \cdots
\label{2.15}
\\
\beta_0 &=& \frac{11}{3} \, C_A - \frac{4}{3} \, T_f \,n_f,
\label{2.16}
\\
\beta_1 &=& \frac{34}{3} C_A^2 - 4 C_F T_f n_f
- \frac{20}{3} C_A T_f n_f .
\label{2.17}
\end{eqnarray}
Further $\delta(\alpha,g)$ is the renormalization group
function which controls the variation of the OMEs
under the gauge constant $\alpha$.
Choosing the general covariant gauge one obtains
in  QCD the following result
\begin{equation}
\delta(\alpha,g) = - \alpha z_\alpha \frac{g^2}{16 \pi^2}  +
\cdots ,
\label{2.18}
\end{equation}
where $z_\alpha$ is given by
\begin{equation}
z_\alpha = \left(-\frac{10}{3} - (1- \alpha) \right) C_A
+ \frac{8}{3} T_f .
\label{smallzalpha}
\end{equation}
Furthermore the colour factors of SU(N) are defined by
$C_A = N$, $C_F = (N^2-1)/(2N)$, $T_f = 1/2$
and $n_f$ stands for the number of light flavours.
The anomalous dimensions are given by the series expansion
\begin{equation}
\gamma_{k,ij}^{m} = \gamma_{k,ij}^{(0),m} \frac{g^2}{16 \pi^2} +
\gamma_{k,ij}^{(1),m} \left(\frac{g^2}{16\pi^2}\right)^2 + \cdots
\label{o2.16}
\end{equation}
Notice that for the subsequent part of this paper we do not need
higher order terms in $\beta(g)$, $\delta(\alpha, g)$
and $\gamma_{k,ij}^{m}$.
As an alternative to using the renormalized OMEs the
anomalous dimensions can also be derived from the operator
renormalization constants $Z_{k,ij}^{m}$ which relate
the bare operators $\hat{R}_{i,k}$ to the renormalized operators
$R_{i,k}$ \footnote{Notice that in the subsequent part of the paper the
unrenormalized quantities will be indicated by a hat.}.
The renormalization of the non-singlet operator
proceeds as
\begin{eqnarray}
\hat{R}_{NS,q}^{\mu_1 \cdots \mu_m} (z)
&=&
Z_{NS,qq}^{m}(\varepsilon,g)
R_{NS,q}^{\mu_1 \cdots \mu_m} (z).
\label{o2.17}
\end{eqnarray}
Since the singlet operators in (\ref{2.6})
and (\ref{2.7}) mix among each other the operator
renormalization constant becomes a matrix and we
have
\begin{eqnarray}
\hat{R}_{S,i}^{\mu_1 \cdots \mu_m} (z)
&=&
Z_{S,ij}^{m}(\varepsilon,g)
R_{S,j}^{\mu_1 \cdots \mu_m} (z).
\label{o2.18}
\end{eqnarray}
Now the anomalous dimensions also can be obtained from
\begin{eqnarray}
\gamma_{NS,qq}^{m} &=& \beta (g,\varepsilon)  \,
Z_{NS,qq}^{-1} \, \frac{d }{dg} Z_{NS,qq},
\no
\gamma_{S,ij}^{m}  &=& \beta (g,\varepsilon)\, (Z_S^{-1})_{il}\,
\frac{d Z_{S,lj}}{dg},
\label{2.19}
\end{eqnarray}
where
\begin{equation}
\beta(g, \varepsilon) = \frac{1}{2} \varepsilon g + \beta (g).
\label{2.20}
\end{equation}
Here $\varepsilon = n-4$ indicates that we will use $n$-dimensional
regularization to regularize the ultraviolet
singularities occurring in $Z_{k,ij}$ which are
represented by pole terms of the type $1/\varepsilon^p$.
The computation of the OMEs proceeds in the
following way. First one adds the operators
(\ref{2.5})--(\ref{2.7})
to the QCD effective lagrangian by multiplying them by sources
$J_{\mu_1 \cdots \mu_m}(z)$. The calculation simplifies
considerably if the sources are chosen to be equal to
$J_{\mu_1 \cdots \mu_m}(z) = \Delta_{\mu_1}
\cdots  \Delta_{\mu_m}$ with $\Delta^2 = 0$. In this way one eliminates
the trace terms on the right-hand side of eq.~(\ref{2.11}).
The Feynman rules for the quark and gluon operator vertices
are given in Appendix A.
Starting from the bare lagrangian, which is expressed in the
bare coupling constant and bare fields and operators, one
obtains the following general form for the unrenormalized OMEs.
For the non-singlet OME we have
\begin{eqnarray}
\hat{A}_{NS,qq} &=& 1 +
\left(\frac{\hat{g}^2}{16 \pi^2}\right) S_\e
\left(\frac{-p^2}{\mu^2}\right)^{\varepsilon/2} \left[\gamma_{NS, qq}^{(0)}
\frac{1}{\varepsilon} + a_{NS,qq}^{(1)}
+ \varepsilon \, b_{NS,qq}^{(1)}
\right]
\nonumber \\
&&
+
 \left(\frac{\hat{g}^2}{16 \pi^2}\right)^2 S^2_\e
\left(\frac{-p^2}{\mu^2}\right)^{\varepsilon}
\left[ \left\{ \frac{1}{2} (\gamma_{NS,qq}^{(0)})^2 -
\beta_0 \gamma_{NS,qq}^{(0)}
      \right\} \frac{1}{\varepsilon^2}
\right.
\nonumber \\
&&+
\left.
\left\{\frac{1}{2} \gamma_{NS,qq}^{(1)} - 2 \beta_0 \,
a_{NS,qq}^{(1)} + \gamma_{NS,qq}^{(0)} a_{NS,qq}^{(1)}
- \hat{\alpha}
\left(
\frac{d\,a_{NS,qq}^{(1)}}{d \hat{\alpha}}
\right)
z_\alpha  \right\} \frac{1}{\epsilon}
\right.
\nonumber \\
&&
\left.
+ a_{NS,qq}^{(2)} \right]_{\hat{\alpha} = 1}.
\label{2.21}
\end{eqnarray}
Here $S_\e$ is a factor which originates from $n$-dimensional
regularization. It is defined by
\begin{equation}
S_\e = e^{\frac{\e}{2} (\gamma_e - \ln 4 \pi)}.
\label{2.22}
\end{equation}
The singlet quark OME can be written as
\begin{equation}
\hat{A}_{S,qq} = \hat{A}_{NS,qq} + \hat{A}_{PS,qq},
\label{2.23}
\end{equation}
where the pure singlet (PS) part is given by
\begin{eqnarray}
\hat{A}_{PS,qq} &=&
\left(\frac{\hat{g}^2}{16 \pi^2}\right)^2 S_{\e}^2
\left(\frac{-p^2}{\mu^2}\right)^{\varepsilon}
\left[ \frac{1}{2} \gamma_{S,qg}^{(0)} \gamma_{S,gq}^{(0)}
\frac{1}{\varepsilon^2}
\right.
\nonumber \\
&&
\left.
\phantom{
\left(\frac{\hat{g}^2}{16 \pi^2}\right)
\left(\frac{-p^2}{\mu^2}\right)^{\varepsilon}
} \;
+ \left\{\frac{1}{2} \gamma_{PS,qq}^{(1)} +
\gamma_{S,qg}^{(0)} a_{S,gq}^{(1)} \right\} \frac{1}{\varepsilon}
+a_{PS,qq}^{(2)} \right],
\label{2.24}
\end{eqnarray}
so that $\gamma_{S,qq}^{(1)} =
\gamma_{NS,qq}^{(1)} + \gamma_{PS,qq}^{(1)}$.

\noindent
The other OMEs can be expressed in the renormalization group
coefficients as
\begin{eqnarray}
\hat{A}_{S,qg} & = &
\left(\frac{\hat{g}^2}{16 \pi^2}\right)
S_\e
\left(\frac{-p^2}{\mu^2}\right)^{\varepsilon/2} \left[
\gamma_{S,qg}^{(0)} \frac{1}{\varepsilon} + a_{S,qg}^{(1)}
+ \e \, b_{S,qg}^{(1)}
\right]
\nonumber \\
&&
+
\left(\frac{\hat{g}^2}{16 \pi^2}\right)^2
S_\e^2
\left(\frac{-p^2}{\mu^2}\right)^{\varepsilon}  \left[
\left\{
\frac{1}{2} \,
\gamma_{S,qg}^{(0)} \,  \left(\gamma_{S,qq}^{(0)}+\gamma_{S,gg}^{(0)}\right)
- \beta_0 \, \gamma_{S,qg}^{(0)}
\right\} \frac{1}{\varepsilon^2}
\right.
\nonumber \\
&&
+
\left\{
\frac{1}{2}\, \gamma_{S,qg}^{(1)} -
2 \beta_0 \, a_{S,qg}^{(1)} +
\gamma_{S,qg}^{(0)}\,  a_{S,gg}^{(1)} +
\gamma_{S,qq}^{(0)} \, a_{S,qg}^{(1)}
- \hat{\alpha}
\left( \frac{d a_{s,qg}^{(1)}}{d \hat{\alpha}}
\right)
 z_\alpha
\right\}  \frac{1}{\varepsilon}
\nonumber\\
&&
\left.
+ a_{S,qg}^{(2)}
\right]_{\hat{\alpha} = 1},
\label{2.25}
\end{eqnarray}
\begin{eqnarray}
\hat{A}_{S,gq} & = &
\left(\frac{\hat{g}^2}{16 \pi^2}\right) S_\e
\left(\frac{-p^2}{\mu^2}\right)^{\varepsilon/2} \left[
\gamma_{S,gq}^{(0)}\frac{1}{\varepsilon} + a_{S,gq}^{(1)}
+\e \, b_{S, gq}^{(1)}
\right]
\nonumber \\
&&
+
\left(\frac{\hat{g}^2}{16 \pi^2}\right)^2 S^2_\e
\left(\frac{-p^2}{\mu^2}\right)^{\varepsilon}  \left[
\left\{
\frac{1}{2} \gamma_{S,gq}^{(0)} \,
\left(\gamma_{S,qq}^{(0)}+\gamma_{S,gg}^{(0)}\right)
- \beta_0 \, \gamma_{S,gq}^{(0)}
\right\} \frac{1}{\varepsilon^2}
\right.
\nonumber \\
&&
+
\left\{
\frac{1}{2} \gamma_{S,gq}^{(1)} -
2 \beta_0 \, a_{S,gq}^{(1)} +
\gamma_{S,gq}^{(0)}\,  a_{S,qq}^{(1)} +
\gamma_{S,gg}^{(0)} \, a_{S,gq}^{(1)}
- \hat{\alpha}
\left(
\frac{d a_{s,gq}^{(1)}}{d \hat{\alpha}}
\right)
 z_\alpha
\right\}  \frac{1}{\varepsilon}
\nonumber\\
&&
\left.
+ a_{S,gq}^{(2)}
\right]_{\hat{\alpha} = 1},
\label{2.26}
\end{eqnarray}
\begin{eqnarray}
\hat{A}_{S,gg} & = &
1+\left(\frac{\hat{g}^2}{16 \pi^2}\right) S_\e
\left(\frac{-p^2}{\mu^2}\right)^{\varepsilon/2} \left[
\gamma_{S,gg}^{(0)}\frac{1}{\varepsilon} + a_{S,gg}^{(1)}
+ \e \, b_{S,gg}^{(1)}
\right]
\nonumber \\
&&
+
\left(\frac{\hat{g}^2}{16 \pi^2}\right)^2 S^2_\e
\left(\frac{-p^2}{\mu^2}\right)^{\varepsilon}  \left[
\left\{
\frac{1}{2} \,\left(\gamma_{S,gg}^{(0)}\right)^2
+\frac{1}{2} \, \gamma_{S,gq}^{(0)} \, \gamma_{S,qg}^{(0)}
- \beta_0 \, \gamma_{S,gg}^{(0)}
\right\} \frac{1}{\varepsilon^2}
\right.
\nonumber \\
&&
+
\left\{
\frac{1}{2} \, \gamma_{S,gg}^{(1)} -
2 \beta_0 \, a_{S,gg}^{(1)}
+
\gamma_{S,gg}^{(0)}\,  a_{S,gg}^{(1)} +
\gamma_{S,gq}^{(0)} \, a_{S,qg}^{(1)}
- \hat{\alpha}
 \left(
 \frac{d a_{S,gg}^{(1)}}{d \hat{\alpha}}
\right)
\, z_\alpha
\right\}  \frac{1}{\varepsilon}
\nonumber\\
&&
\left.
+ a_{S,gg}^{(2)}
\right]_{\hat{\alpha}=1}.
\label{2.27}
\end{eqnarray}
Notice that in the above we have suppressed the Mellin
index $m$.
The expressions have been written in such a way that
the anomalous dimensions take their values in the
\MSbs.
Furthermore we have in lowest order the identity
$\gamma_{S,qq}^{(0)} = \gamma_{NS,qq}^{(0)}$.
The above form of the unrenormalized OMEs
$\hat{A}_{k,ij}$ follows from the property that the
renormalized OMEs $A_{k,ij}$ satisfy the Callan Symanzik
equations  (\ref{2.13}), (\ref{2.14}).
These equations can be solved order by order in perturbation theory
which provides us with the expressions presented at the end of
this section. We only have to show that the latter follow from
the renormalization of the OMEs in (\ref{2.21})--(\ref{2.27}).

The renormalization of the OMEs proceeds
as follows. First replace the bare coupling constant
$\hat{g}$ by the renormalized one $g(\mu)=g$.
Up to order $\hat{g}^4$ it is sufficient
to substitute in the above OMEs
\begin{equation}
\hat{g} = g \left(1+\frac{g^2}{16 \pi^2}
\,  \beta_0  \, S_\varepsilon \,
\frac{1}{\varepsilon}\right),
\end{equation}
where $\beta_0$ is given by (\ref{2.16}).
Next one has to perform gauge constant renormalization.
Notice that in the next section we will
calculate the one-loop OMEs in a general
covariant gauge. The Feynman propagator in this gauge is
given by
\begin{equation}
D_{\mu \nu}(k) = \frac{1}{k^2+i \epsilon}
\left(-g_{\mu \nu} + (1-\alpha)
\frac{k_\mu k_\nu}{k^2 + i \epsilon}
\right),
\end{equation}
where $\alpha$ is the gauge constant.
The two-loop OMEs are computed in the Feynman gauge so that
we have put in eqs.~(\ref{2.21})--(\ref{2.27})
 $\hat{\alpha} = 1$.

\noindent
Since the quarks and gluons are massless one
has to put the external
momenta $p$ of the Feynman graphs off-shell.
This implies that the OMEs are no longer
S-matrix elements and they become
gauge ($\alpha$) dependent.
Therefore we also have to perform gauge constant
renormalization which proceeds as follows.
Replace the bare gauge constant $\hat{\alpha}$
by the renormalized one.

\begin{equation}
\hat{\alpha} = Z_\alpha \alpha .
\end{equation}
In the covariant gauge one has the property
$Z_\alpha = Z_A$ where $Z_A$ is the gluon
field renormalization constant.
Hence $Z_\alpha$ is given by
\begin{equation}
Z_\alpha= 1 + \frac{g^2}{16 \pi^2} z_\alpha \frac{1}{\varepsilon}
\label{zalpha},
\end{equation}
where $z_\alpha$ is given in (\ref{smallzalpha}).
After these two renormalizations the only ultraviolet
divergences left in the OMEs are removed by operator
renormalization.
Choosing the \MSbs\ the operator
renormalization constants are given by (see (\ref{o2.17}), (\ref{o2.18}))
\begin{eqnarray}
Z_{NS,qq} &=&
1 + \left( \frac{g^2}{16 \pi^2} \right) \, S_\varepsilon
\, \left[\frac{1}{\varepsilon} \gamma^{(0)}_{NS,qq} \right]
\no
&&
+ \left( \frac{g^2}{16 \pi^2} \right)^2 S^2_\varepsilon
\left[\left\{\frac{1}{2} \left( \gamma^{(0)}_{NS,qq} \right)^2 +
\beta_0 \gamma^{(0)}_{NS,qq} \right\} \frac{1}{\varepsilon^2} +
\frac{1}{2 \varepsilon} \gamma^{(1)}_{NS,qq} \right],
\label{2.33}
\\
Z_{S,qq} &=& Z_{NS,qq} + Z_{PS,qq},
\label{2.34}
\end{eqnarray}
with
\begin{equation}
Z_{PS,qq} = \left( \frac{g^2}{16 \pi^2} \right)^2 S^2_\varepsilon
\left[\left\{\frac{1}{2} \gamma^{(0)}_{S,qg} \gamma^{(0)}_{S,gq}
     \right\} \frac{1}{\varepsilon^2} +
\frac{1}{2 \varepsilon} \gamma^{(1)}_{PS,qq} \right],
\label{2.35}
\end{equation}
\begin{eqnarray}
Z_{S,qg} &=&
\phantom{+}
\left( \frac{g^2}{16 \pi^2} \right) \, S_\varepsilon \,
\left[ \frac{1}{\varepsilon} \gamma_{S,qg}^{(0)} \right]
\label{2.36}
\\
&&
+
\left( \frac{g^2}{16 \pi^2} \right)^2 \, S_\varepsilon^2 \,
\left[
\left\{
\frac{1}{2} \gamma_{S,qg}^{(0)} \left(\gamma_{S,gg}^{(0)} +
                                     \gamma_{S,qq}^{(0)}
                              \right)
+
\beta_0 \, \gamma_{S,qg}^{(0)}
\right\} \, \frac{1}{\varepsilon^2} +
\frac{1}{2 \varepsilon} \gamma_{S,qg}^{(1)}
\right],
\nonumber
\\
Z_{S,gq} &=&
\phantom{+}
\left( \frac{g^2}{16 \pi^2} \right) \, S_\varepsilon \,
\left[ \frac{1}{\varepsilon} \gamma_{S,gq}^{(0)} \right]
\label{2.37}
\\
&&
+
\left( \frac{g^2}{16 \pi^2} \right)^2 \, S_\varepsilon^2 \,
\left[
\left\{
\frac{1}{2}\, \gamma_{S,gq}^{(0)} \left(\gamma_{S,gg}^{(0)} +
                                     \gamma_{S,qq}^{(0)}
                              \right)
+
\beta_0 \, \gamma_{S,gq}^{(0)}
\right\} \, \frac{1}{\varepsilon^2} +
\frac{1}{2 \varepsilon} \gamma_{S,gq}^{(1)}
\right],
\nonumber
\\
Z_{S,gg} &=&
1 +
\left( \frac{g^2}{16 \pi^2} \right) \, S_\varepsilon \,
\left[ \frac{1}{\varepsilon} \gamma_{S,gg}^{(0)} \right]
\nonumber \\
&&
\phantom{1}
+
\left( \frac{g^2}{16 \pi^2} \right)^2 \, S_\varepsilon^2 \,
\left[
\left\{
\frac{1}{2} \left(\gamma_{S,gg}^{(0)} \right)^2
 + \frac{1}{2}\, \gamma_{S,gq}^{(0)} \gamma_{S,qg}^{(0)}
 + \beta_0 \, \gamma_{S,gg}^{(0)}
\right\} \, \frac{1}{\varepsilon^2}
\right.
\nonumber \\
&&
\phantom{
\left( \frac{g^2}{16 \pi^2} \right)^2 \, S_\varepsilon^2 \,
\left[
\left\{
\right. \right.
}
\left.
 +
\frac{1}{2 \varepsilon} \gamma_{S,gg}^{(1)}
\right],
\label{2.38}
\end{eqnarray}

Notice that the anomalous dimensions
$\gamma_{k,ij}^{(l)}$ $(k = $NS,S, $l =0,1)$ are gauge
independent so that $Z_{k,ij}$ have to be gauge independent too.
The renormalized operator matrix elements are derived from
\begin{equation}
A_{NS,qq}(p^2,\mu^2,g,\alpha) =
Z^{-1}_{NS,qq}(g^2,\varepsilon)
\hat{A}_{NS,qq}(p^2,\mu^2,\hat{g},\hat{\alpha})
\vert_{\hat{g} \rightarrow Z_gg; \;\;
       \hat{\alpha} \rightarrow Z_\alpha \alpha}
\label{2.39}
\end{equation}
and
\begin{equation}
A_{S,ij}(p^2,\mu^2,g,\alpha) =
(Z^{-1}_S)_{il}(g^2,\varepsilon)
\hat{A}_{S,lj}(p^2,\mu^2,\hat{g},\hat{\alpha})
\vert_{g \rightarrow Z_gg; \;\;
       \hat{\alpha} \rightarrow Z_\alpha \alpha}
\label{2.40}
\end{equation}
with the results
\begin{eqnarray}
A_{NS,qq} &=&
1 + \frac{g^2}{16 \pi^2}
\left[
\frac{1}{2} \gamma_{NS,qq}^{(0)}
\ln \left(- \frac{p^2}{\mu^2} \right)
+a_{NS,qq}^{(1)} \right]
\nonumber \\
&&
+\left(
\frac{g^2}{16 \pi^2}
\right)^2
\left[
\left\{
\frac{1}{8} \left( \gamma_{NS,qq}^{(0)}\right)^2 -
\frac{1}{4} \beta_0 \, \gamma_{NS,qq}^{(0)}
\right\} \, \ln^2 \left(-\frac{p^2}{\mu^2} \right)
\right.
\nonumber \\
&&
+
\left\{
\frac{1}{2} \gamma_{NS,qq}^{(1)} -
\beta_0 a_{NS,qq}^{(1)} +
\frac{1}{2} \gamma_{NS,qq}^{(0)} a_{NS,qq}^{(1)}
- \frac{1}{2} \alpha
\left( \frac{d a_{NS,qq}^{(1)}}{d \alpha} \right) z_\alpha
\right\}
 \, \ln \left(-\frac{p^2}{\mu^2} \right)
\nonumber \\
&&
\times \left.
a_{NS,qq}^{(2)} + 2 \beta_0 \, b_{NS,qq}^{(1)} -
\gamma_{NS,qq}^{(0)} b_{NS,qq}^{(1)} +
\alpha \left( \frac{d}{d \alpha} b_{NS,qq}^{(1)}
 \right) z_\alpha
\right]_{\alpha =1},
\label{2.41}
\end{eqnarray}
\begin{eqnarray}
A_{S,qq} &=& A_{NS,qq} + A_{PS,qq},
\nonumber \\
A_{PS,qq} &=&
\gsgs
\left[
\frac{1}{8} \gqg \ggq \lnpp
\right.
\no
&&
\phantom{ \gsgs \; }
+
\left\{ \frac{1}{2} \gpsqq + \frac{1}{2} \gqg \agq \right\} \lnp
\no
&&
\phantom{ \gsgs \; }
\left.
+ \apsqq - \gqg \bgq
\right],
\\
A_{S,qg} &=& \gs
\left[
\frac{1}{2} \gqg \lnp + \aqg
\right]
\no
&+&
\gsgs
\left[
\left\{
\frac{1}{8} \gqg \left(
\gnsqq + \GGG \right) - \frac{1}{4} \beta_0 \, \gqg
\right\}
\lnpp
\right.
\no
&&
\phantom{
\gsgs \;
}
+ \left\{ \frac{1}{2} \Gqg -
\beta_0 \aqg + \frac{1}{2} \gqg \agg +
\frac{1}{2} \gnsqq \aqg
\right.
\no
&&
\left.
\phantom{ \gsgs \; }
-\frac{1}{2} \alpha \left( \frac{d}{d \alpha} \, \aqg \right) z_\alpha
\right\} \, \lnp
\no
&&
\phantom{ \gsgs \; }
+
\Aqg + 2 \beta_0 \aqg - \gnsqq \bqg - \gqg \bgg
\no
&&
\left.
\phantom{ \gsgs \; }
+
\alpha
\left(
\frac{d}{d \alpha} \bqg \right) z_\alpha
\right]_{\alpha=1},
\label{2.42}
\\
A_{S,gq} &=& \gs
\left[
\frac{1}{2} \ggq \lnp + \agq
\right]
\no
&+&
\gsgs
\left[
\left\{
\frac{1}{8} \ggq \left(
\gnsqq + \GGG \right) - \frac{1}{4} \beta_0 \, \ggq
\right\}
\lnpp
\right.
\no
&&
\phantom{
\gsgs \;
}
+ \left\{ \frac{1}{2} \Ggq -
\beta_0 \agq + \frac{1}{2} \ggq \ansqq +
\frac{1}{2} \GGG \agq
\right.
\no
&&
\left.
\phantom{ \gsgs \; }
-\frac{1}{2} \alpha \left( \frac{d}{d \alpha} \, \agq \right) z_\alpha
\right\} \, \lnp
\no
&&
\phantom{ \gsgs \; }
+
\Agq + 2 \beta_0 \agq - \ggq b_{S,qq}^{(1)} - \GGG \bgq
\no
&&
\left.
\phantom{ \gsgs \; }
+
\alpha
\left(
\frac{d}{d \alpha} \bgq \right) z_\alpha
\right]_{\alpha = 1},
\label{2.43}
\\
A_{S,gg} &=&
1 + \gs
\left[ \frac{1}{2} \GGG \lnp + \agg \right]
\no
&+&
\gsgs
\left[
\left\{
\frac{1}{8} \left(\GGG \right)^2
+ \ggq \gqg  - \frac{1}{4} \beta_0 \, \GGG
\right\}
\lnpp
\right.
\no
&&
\phantom{
\gsgs \;
}
+ \left\{ \frac{1}{2} \Ggg -
\beta_0 \agg + \frac{1}{2} \GGG \agg +
\frac{1}{2} \ggq \aqg
\right.
\no
&&
\left.
\phantom{ \gsgs \; }
-\frac{1}{2} \alpha \left( \frac{d}{d \alpha} \, \agg \right) z_\alpha
\right\} \, \lnp
\no
&&
\phantom{ \gsgs \; }
+
\Agg + 2 \beta_0 \bgg - \GGG \bgg - \ggq \bqg
\no
&&
\left.
\phantom{ \gsgs \; }
+
\alpha
\left(
\frac{d}{d \alpha} \bgg \right) z_\alpha
\right]_{\alpha = 1}.
\label{2.44}
\end{eqnarray}
The above renormalized OMEs satisfy the Callan Symanzik
equations in (\ref{2.13}),  (\ref{2.14}) which proves that
the ansatz for the unrenormalized OMEs in
(\ref{2.21})--(\ref{2.27}) is correct.
This is also corroborated by the expressions for the operator
renormalization constants $Z_{k,ij}$ in
(\ref{2.33})--(\ref{2.38}) which after insertion in
eqs.~(\ref{2.19}), (\ref{2.20}) provides us with the
anomalous dimensions in (\ref{o2.16}).

The above renormalization procedure was originally introduced by
Dyson \cite{X18}. There exists an alternative
possibility invented by Bogoliubov, Parasiuk, Hepp and Zimmermann
(BPHZ) \cite{X19}. In the latter one renormalizes each
Feynman graph independently using the counter-term method.
These counter-terms appear in the effective lagrangian which is
expressed into the renormalized (coupling- and gauge-) constants,
fields and operators. The BPHZ-method has been used in the
literature \cite{X14}--\cite{X17}
to derive the anomalous dimensions of the spin
averaged operators $O^{\mu_1 \cdots \mu_m}$ (i=NS,S) in
(\ref{2.1}).  The advantage of this method is that the gauge
dependent terms given by
$\alpha \left(\frac{d}{d \alpha} \, a_{k,ij}^{(1)}\right)
z_\alpha$ in
(\ref{2.21})--(\ref{2.27}) are automatically subtracted. We will come
back to this method at the end of section 3.
The reason for the
algebraic exercise given above can be explained as follows. Since the
lowest order coefficients $\gamma_{k,ij}^{(0)}$ and
$a_{k,ij}^{(1)}$ can be very easily determined
from the one-loop OMEs
one
immediately
can
predict the double pole terms in the unrenormalized OMEs
(\ref{2.21})--(\ref{2.26}). The coefficient of
the single pole term can be
also computed except for the second order anomalous dimensions
$\gamma _{k,ij}^{(1)}$. By equating the predicted form of the
two-loop OMEs in (\ref{2.21})--(\ref{2.26}) to the explicitly
computed result in the next section one immediately can
infer the results for $\gamma_{k,ij}^{(1)}$.

\section{Calculation of the order $\alpha_s^2$ contribution to the
spin splitting functions}
In this section we first give an outline of the procedure of the
calculation of the OMEs defined in (\ref{2.11}).
Then we present the analytical result for the OMEs and
extract from them the splitting functions (anomalous dimensions).

The calculation of the OMEs proceeds as follows. Using the
operator vertices in Appendix A and applying the standard
Feynman rules we have computed the connected Green's functions,
which are given by the graphs in Figs.~1-6, up to two-loop order.
The latter also involves the calculation of the
diagrams which contain the
self energies of the quark and the gluon in the external legs.
These diagrams are not explicitly drawn in the figures but
are included in our calculation.
The computation of the
one-loop graphs has been done in the general covariant gauge
because one has to renormalize the gauge constant $\alpha$ even if one
chooses the Feynman gauge $\alpha = 1$. The two-loop graphs have been
calculated in the Feynman gauge which is sufficient to that order.
The OMEs are then obtained by multiplying the connected Green`s function by
the inverse of the external quark and gluon propagators. Since the external
momenta are put off shell only ultraviolet divergences appear in the
OMEs which are regularized by using the method of $n$-dimensional
regularization. This implies that we have to find a suitable
prescription for the $\gamma_5$-matrix which appears in the
quark operators $R_{k,q}$ for $k = NS$ (\ref{2.5}) and $k=S$ (\ref{2.6}).
Here we will adopt the reading point method as explained in
\cite{X20}.

 As reading point we will choose the operator vertex which implies that the 
 results for the unrenormalized OMEs are the same as those obtained by using the prescription of 
 `t Hooft and Veltman \cite{X21},  (see also \cite{X22}, \cite{X23}).
 This method is characterized by the phenomenon that the non-singlet axial vector operator 
 $R_{NS,q}^{(1)}$  (\ref{2.5}) gets renormalized in spite of the fact that it is conserved. 
 This effect has to be undone by introducing an additional renormalization
 constant \cite{X23}.
However, for continuity this procedure has to be extended to
higher spin non-singlet operators $R_{NS,q}^{m}$ (\ref{2.5})
$(m>1)$ otherwise the anomalous dimension of $R_{NS,q}^{m}$
will become unequal to
the anomalous dimension
of the spin averaged non-singlet operator $O_{NS,q}^{m}$
(see (\ref{2.1})).
Notice that the same procedure has to be also carried out for 
the singlet operators $R_{S,q}^{m}$ (\ref{2.6}).
Summarizing our procedure we first calculate the unrenormalized OMEs and determine the 
splitting functions. Then we subsequently perform an additional finite renormalization to undo the 
unwanted effects of the prescription for the $\gamma_5$-matrix.
However, this procedure can be avoided for the non-singlet OME as we will show below.

As has been already mentioned in section 2 the Feynman rules for
the operator vertices in Appendix A have been derived multiplying
the operators $R_{k,i}^{\mu_1 \cdots \mu_m}$ by the sources
$J_{\mu_1 \cdots \mu_m} = \Delta_{\mu_1} \cdots \Delta_{\mu_m}$
with $\Delta^2 = 0$. To simplify further we can choose
$s=p$, where $s$ is the spin vector in
(\ref{2.4}), without any loss of essential information.
The operator matrix elements $\hat{A}_{k,ij}^{m}$
(\ref{2.21})--(\ref{2.27}) are then given by
\begin{equation}
\hat{A}_{k,iq}^{m}(p^2,\mu^2,g,\varepsilon) (\Delta p)^m
= \frac{1}{4} \,
Tr \left\{ \hat{G}_{k,iq}^{m}(p,\Delta,\mu^2,g,\varepsilon)
\gamma_5
\ps
  \right\},
\label{3.1}
\end{equation}
with $k = NS,S$ and $i = q,g$ and
\begin{equation}
\hat{A}_{S,ig}^{m}(p^2,\mu^2,g,\varepsilon) (\Delta p)^m
= \frac{1}{2\, \Delta p} \,
\varepsilon_{\mu \nu \lambda \sigma} \Delta^\lambda
p^\sigma \,
\hat{G}_{S,ig}^{m \mu \nu}(p,\Delta, \mu^2,g,\varepsilon).
\label{3.2}
\end{equation}
Here $\hat{G}_{k,ij}^{m}$ stand for the unrenormalized Green`s
functions which are multiplied by the inverse of the external
quark and gluon propagators.

We will now give a short outline of the calculation of
$\hat{A}_{k,ij}^{m}$.
Let us first start with the non-singlet OME
$\hat{A}^{m}_{NS,qq}$.
The Green`s function $\hat{G}_{NS,qq}^{m}$, which is determined
by the one-loop graphs in Fig.~1a,b
and by the two-loop graphs in Fig.~2,
consists out of Feynman integrals where the numerators are
given by a string of $\gamma $-matrices. One of the $\gamma$-matrices 
represents the $\gamma_5$-matrix. The latter is then
anticommuted with the other $\gamma$-matrices until it appears on
the right side of the string next to the $\gamma_5$ in
(\ref{3.1}). Then we set $\gamma_5^2 = 1$ and simplify the
trace by contracting over dummy Lorentz-indices. Finally we perform
the trace in (\ref{3.1}). In this way one obtains the identity
\begin{equation}
\hat{A}_{NS,qq}^{m}(p^2,\mu^2,\varepsilon) =
\left(\hat{A}_{NS,qq}^{m}(p^2,\mu^2,\varepsilon)
\right)_{\rm spin-averaged},
\label{3.3}
\end{equation}
without any additional renormalization constant.
Notice that the calculation of $A_{NS,qq}^{m}$
(spin averaged OME) has been already done in the literature
so that it will not be repeated here.
Except for the non-singlet operator
$\hat{A}_{NS,qq}^{m}$ the remaining
spin OMEs differ from their
spin averaged analogues. Since we need the one-loop OMEs as
presented in Fig.~1, for the renormalization of the two-loop
OMEs given by Figs.~2-6  we have to calculate the
former ones up to the non-pole term $a_{k,ij}$ defined in
eqs.~(\ref{2.21})--(\ref{2.27}).
The one-loop terms $b_{k,ij}$ which are proportional to
$\e = n-4$,
do not play any role
in the determination of the anomalous dimension
and they will not be presented in this paper.

Starting with the one-loop contribution to
$\hat{A}_{S,gq}^{m}$ (Fig.~1c) we have to
perform tensorial reduction of the tensor integrals appearing in
$\hat{G}_{S,gq}^{m}$. These tensor integrals arise
because the integration momentum $q_\mu$
appears in the numerators
of the integrand. Examples of such one-loop integrals are
given by eqs.~(\ref{B.1})--(\ref{B.3}).
The result will be that $\hat{G}_{S,gq}^{m}$ gets terms of the
form $\e^{\alpha \beta \lambda \sigma} \Delta_\sigma
p_\lambda  \gamma_\alpha \ps \gamma_\beta$,
$\e^{\alpha \beta \lambda \sigma} \Delta_\sigma
  \gamma_\alpha
\gamma_\lambda
\gamma_\beta
$,
$\e^{\alpha \beta \lambda \sigma}
p_\lambda \Delta_\sigma  \gamma_\alpha \Ds \gamma_\beta$,
where the Levi-Civita tensor $\e^{\alpha \beta \lambda \sigma}$
originates from the two-gluon operator vertex in (\ref{Atwogluon}).
The trace in (\ref{3.1}) provides us with a second
 Levi-Civita tensor so that we have to contract over
two and three dummy Lorentz-indices. The contraction
has to be performed in 4 dimensions since the operator vertices
have a unique meaning in 4 dimensions only.
Next we discuss the calculation of
the one-loop contribution to
$\hat{A}_{S,qg}^{m}$ (\ref{3.2}) (Fig.~1d,e).
To this OME we apply the reading point method
\cite{X20}
and put the $\gamma_5$ on the right hand side of the trace from the
start. In this way we reproduce the ABJ-anomaly which
can be traced back to the triangular fermion loop in Fig.~1d.
Notice that Fig.~1e leads to a zero result because the external
momentum $p$ appears twice in the  Levi-Civita tensor.
The Green`s function $\hat{G}_{S,qg}^{m\mu \nu}$ will then become
proportional to $\e^{\mu \nu \lambda \sigma} \Delta_\lambda p_\sigma$.
The latter will be contracted with the  Levi-Civita tensor in
 (\ref{3.2}) where the contraction is performed in 4 dimensions.
The one-loop graphs contributing to $\hat{A}_{S,gg}^{m}$
are presented in Figs.~1f, g.
Because of the  Levi-Civita tensor
coming from the two-gluon operator vector in (\ref{Atwogluon})
$\hat{G}_{S,gg}^{m,\mu \nu}$ will,
after tensorial reduction, become
proportional to $\e^{\mu \nu \lambda \sigma} \Delta_\lambda p_\sigma$.
Like in the case of $\hat{A}_{S,qg}^{m}$ the contraction with the
Levi-Civita tensor in (\ref{3.2}) has to take place in 4 dimensions.

Before we proceed with the two-loop graphs we want to emphasize
that first the tensorial reduction has to be made before
one can perform the contraction between the two Levi-Civita tensors.
Both operations do not commute and lead to different results for
the OMEs. This holds for the one as well as two-loop calculation.
If one contracts the Levi-Civita tensors in $n$ dimensions both
operations commute. However, then the Lorentz indices of
the operator vertices in Appendix A have to be generalized to
$n$ dimensions which is a non-unique procedure.

\noindent
The calculation of the two-loop graphs in Figs.~3-6 proceeds in an
analogous way as in the one-loop case. However, here there arise
some extra complications. First of all we encounter the two-loop
scalar Feynman integrals which have already been performed
in \cite{X24} to calculate the spin averaged
OME $\hat{A}_{S,gg}^{m}$.
To check these integrals and the tensorial reduction algorithm
we have recalculated all spin averaged
anomalous dimensions (splitting functions) and we found
complete agreement with the results published in the
literature \cite{X14}--\cite{X17}.

\noindent
The second complication shows up in the tensorial reductions of the
two-loop tensor Feynman integrals where the numerator
now reveals the presence of two integration momenta $q_1$ and
$q_2$. A more detailed explanation of how the tensor integrals
are reduced into scalar integrals is presented in Appendix B.
The third complication arises because of the appearance of
a trace of six $\gamma $-matrices out of which two are contracted
with the integration momenta $q_1$ and $q_2$.  Such graphs
(see e.g.~Fig.~3 and Figs.~5.11) are calculated by the following procedure.
First one performs tensorial reduction of the Feynman integrals
as indicated in Appendix B. This will lead to an increase of
the pairs of $\gamma$-matrices having the same Lorentz-index.
Then one can eliminate these pairs using the standard rules for
$\gamma $-algebra in $n$ dimensions. This is possible without
ever touching the $\gamma_5$ matrix because it is put at the right
hand side of the string of $\gamma$-matrices. After this procedure
one ends up with the expression $Tr(\as \bs \cs \ds \gamma_5)$ which
is uniquely defined (irrespective of the $\gamma_5$-scheme).
The same  holds for the other graphs which do not contain fermion loops.
The final result is that all Green`s functions $\hat{G}_{k,ij}^{m}$
get the same form as observed for the one-loop case. Four dimensional
contraction of the two Levi-Civita tensors yields the OMEs
$\hat{A}_{k,ij}^{(m)}$ in (\ref{3.1}), (\ref{3.2}).
Before finishing the technical part of this section we give a comment on
the algebraic manipulation programs which are used to calculate
$\hat{A}_{k,ij}^{m}$.
The matrix elements (including the full
tensorial reduction) were calculated using the package
{\it FeynCalc} \cite{X25} which is written in
{\it Mathematica} \cite{X26}.
The two-loop scalar integrals were performed by using a
program written in FORM \cite{X25} which was called in
{\it FeynCalc}.

If one performs the inverse Mellin transform of the OMEs the
results for the one-loop calculation can be summarized as
follows (see eqs.~(2.21)--(2.27)).
First we have the lowest
order splitting functions which are already known in the literature
\cite{X12,X13}.
\begin{eqnarray}
P_{NS,qq}^{(0)} &=& P_{S,qq}^{(0)} =
C_F \, \left[ 8 \pld - 4 - 4 x + 6 \dx \right],
\label{3.4}
\\
P_{S,qg}^{(0)} &=& T_f \, \left[16 x - 8 \right],
\label{3.5}
\\
P_{S,gq}^{(0)} &=& C_f \, \left[8 - 4 x \right],
\label{3.6}
\\
P_{S,gg}^{(0)} &=& C_A\, \left[8 \pld + 8 - 16 x
+ \frac{22}{3} \dx \right] - T_f \left[ \frac{8}{3} \dx \right],
\label{3.7}
\end{eqnarray}
where the colour factors in SU(N) are given by
$C_F = (N^2-1)/(2 N), C_A = N$ and $T_f = 1/2$ ($N=3$ in QCD).
The non-pole terms $a_{k,ij}^{(1)}$ appearing in
expressions (\ref{2.21})--(\ref{2.27})
read as follows
\begin{eqnarray}
a_{NS,qq}^{(1)} &=& a_{S,qq}^{(1)} =
C_F \,  \left[ -4 \lpld + 2 (1+x) \lmx - 2 \frac{1+x^2}{1-x} \lx
\right.
\no
&& - 4 + 2 x + (1-\alpha) \left(2 - \pld \right) +
\dx \left(7 - 4 \zeta(2) \right),
\label{3.8}
\\
a_{S,qg}^{(1)} &=&
T_f \,  \left[(4-8 x) (\lx + \lmx) - 4 \right],
\label{3.9}
\\
a_{S,gq}^{(1)} &=&
C_F \, \left[
(-4 + 2 x) (\lx + \lmx) + 2 - 4 x
\right],
\label{3.10}
\\
a_{S,gg}^{(1)} &=&
C_A \left[ -4 \lpld + (-4 + 8 x) \lmx +
\left(- \frac{4}{1-x} - 4 + 8 x  \right) \lx
\right.
\no
&&
\left.
-(1-\alpha) \pld + 2 +
\dx \left( \frac{67}{9} - 4 \zeta(2) + (1-\alpha) -
\frac{1}{4} (1-\alpha)^2
\right)
\right]
\no
&&
-T_f \left[ \frac{20}{9} \dx \right].
\label{3.11}
\end{eqnarray}
In the above expressions the distributions
$\left( \frac{\ln^k (1-x)}{1-x} \right)_+$
are defined by
\begin{equation}
\int_0^1 dx \, \left( \frac{\ln^k (1-x)}{1-x} \right)_+ \, f(x)
=
\int_0^1 dx \, \left( \frac{\ln^k (1-x)}{1-x} \right) \, ( f(x) - f(1) ).
\label{3.12}
\end{equation}

In the above one observes that 
in the general covariant gauge only
$a_{k,qq}^{(1)}$ (k=NS,S) and
$a_{S,gg}^{(1)}$ depend on the gauge parameter $\alpha$.
Further notice that in the case of `t Hooft Veltman prescription the expression for
$a_{k,qq}^{(1)}$ (k=NS,S) (\ref{3.8}) only emerges after one has added the renormalization constant
$z_{qq} = -8 C_F (1-x)$. 
\noindent
The two-loop contributions to the unrenormalized
OMEs are given by the inverse Mellin transforms
(see eqs. (\ref{2.21})--(\ref{2.27}))
\begin{eqnarray}
\hat{A}_{PS,qq} &=&
\frac{ g^4}{(16 \pi^2)^2}
\, S^2_\e \,
\left(\frac{-p^2}{\mu^2} \right)^{\e}
C_F \, T_f \, B_{Ff}^{qq},
\label{3.13}
\\
\hat{A}_{S,qg} &=&
\frac{ g^4}{(16 \pi^2)^2}
 \, S^2_\e \, \left(\frac{-p^2}{\mu^2} \right)^{\e}
 \left[
T_f^2\,  B_{ff}^{qg} + C_F \, T_f \, B_{Ff}^{qg} +
C_A\,  T_f \, B_{A f}^{qg}
\right],
\label{3.14}
\\
\hat{A}_{S,gq} &=&
\frac{ g^4}{(16 \pi^2)^2}
\, S^2_\e \, \left(\frac{-p^2}{\mu^2} \right)^{\e}
\left[
      C_A\, C_F \,B_{A F}^{gq} +
C_F\,
T_f\,
B_{Ff}^{gq} +C_F^2\, B_{FF}^{gq}
\right],
\label{3.15}
\\
\hat{A}_{S,gg} &=&
\frac{ g^4}{(16 \pi^2)^2}
 \, S^2_\e \, \left(\frac{-p^2}{\mu^2} \right)^{\e}
\left[ C_A\, T_f \, B_{Af}^{gg}
     + C_F \,T_f \, B_{F f}^{gg}
     + C_A^2 \, B_{AA}^{gg}
\right],
\label{3.16}
\end{eqnarray}

\begin{eqnarray}
B_{Ff}^{qq} &=&
\phantom{\,}
\frac{8}{\e^2} \lc
\frac{}{}
4 \, (1 + x) \ln x + 10 \, (1-x)
\frac{}{} \rc
\no &&
+\frac{8}{\e}
\lc
\frac{}{}
  4 \, (1+x) \li (1-x)
+ 3 \, (1+x) \ln^2 x
+ 4 \, (1+x)
\ln x
\ln (1-x)
\right.
\no && \phantom{\frac{8}{\e} \, }
\left.
+ 10 \, (1-x) \ln (1-x)
+ (7-5\,x) \ln x
- 5\,(1-x)
\frac{}{}
\rc,
\no
B_{ff}^{qg} &=&
\,\frac{64}{\e^2} \lc
\frac{1}{3} \,  (1-2 \, x) \rc
+ \frac{32}{\e}
\lc -\frac{2}{9} \,  (4 - 5 \, x) +
\frac{1}{3} \, ( 1 - 2 \, x) \ln x
\rc,
\no
B_{Ff}^{qg} &=&
\,
\frac{4}{\e^2} \lc
\frac{}{}
 -8 \, (1 - 2 \, x) \lmx + 4 \, (1- 2 \, x) \ln x + 6
\frac{}{}
               \rc
\no &&
+\frac{4}{\e}
\lc
\frac{}{}
- 12 \, (1 - 2 \, x) \, \li (1-x)
+ 4 \, (1 - 2 \, x) \, \zeta( 2)
-6 \, (1 - 2 \, x) \ln^2(1-x)
\right.
\no && \phantom{\frac{4}{\e} \, }
- 8 \,  (1 - 2 \, x) \ln x \lmx
+ 3 \, ( 1 - 2 \, x) \ln^2 x
+ 2 \, (4 \, x + 3 ) \lmx
\no && \phantom{\frac{4}{\e} \, }
\left.
- ( 8 \, x + 5) \ln x
- 12 + 13 \,x
\frac{}{}
\rc,
\no
B_{Af}^{qg} &=&
\,
\frac{16}{\e^2}
\lc
- 2 \, ( 1 - 2 \, x) \lmx
+ 4 \, (1+x) \ln x
+ \frac{1}{3} (25 - 14 \, x)
\right\}
\no &&
+\frac{8}{\e}
\lc \frac{}{}
 12 \, \li (1-x)
+ 2 ( 2 \, x + 1) \li (-x)
- 2 \, (1 - 4 \, x) \, \zeta (2)
\right.
\no && \phantom{\frac{8}{\e} \,}
-3 ( 1 - 2 \, x) \ln^2 (1-x)
+ 2 \, (2 \, x + 1) \ln x \ln (1+x)
+ 8 \, (1+x) \ln x \lmx\
\no  && \phantom{\frac{8}{\e} \,}
\left.
+ (6\, x+5) \ln^2 x
+ \frac{2}{3} (11 \, x + 14) \ln x
+ \frac{1}{3} (73 - 62 \, x) \lmx
- \frac{1}{9} (44 - x)
\rc,
\no
B_{AF}^{gq} &=&
\frac{8}{\e^2} \lc
\frac{1}{3} (25 \, x - 14)
+ 2 \, (2 -x) \lmx
- 2\, (x+4) \ln x
\rc
\no &&
+\frac{4}{\e}
\lc \frac{}{}
- 12 \, x \,\li (1-x)
- 2\, (x+2) \li (-x)
- 2 \, (4-x) \, \zeta(2)
- 2\, (x+2) \ln x \ln (1+x)
\right.
\no && \phantom{+\frac{4}{\e}}
+ 3 \, (2 - x) \ln^2 (1-x)
- (3 \, x + 10) \ln^2 x
- 2 \, (5 x+2) \ln x \lmx
\no && \phantom{+\frac{4}{\e}}
\left.
- \frac{1}{3} (50 - 73 \, x) \lmx
+ \frac{1}{3}\, (17 \, x + 8) \ln x
+ \frac{1}{9} (109 - 119 \, x)
\rc,
\no
B_{Ff}^{gq} &=&
\, \frac{32}{\e^2}
\lc \frac{1}{3} \, (x-2) \rc
+
\frac{16}{\e}\, \lc
\frac{1}{9}\, (10 - 11 x) + \frac{1}{3} (x-2) \, \lmx
+\frac{2}{3}\, (x-2) \ln x
\rc,
\no
B_{FF}^{gq} &=&
\,\frac{4}{\e^2}
\lc
\frac{}{}
 3 \, x - 4 \, ( x - 2 ) \lmx + 2 \, (x-2) \ln x
\frac{}{}
\rc
\no &&+\frac{2}{\e}
\lc
\frac{}{}
  12 \, (x-2) \, \li (1-x)
+ 8 \, (2-x) \, \zeta(2)
+ 6 \, (2-x) \ln^2 (1-x)
+ 3 \, (x-2) \ln^2 x
\right.
\no && \phantom{+\frac{2}{\e}}
\left.
+ 4 \, (x - 2) \ln x \lmx
- 2 \, (4 - 7 \, x) \lmx
+ 7 \, x \ln x
+ 9-14\, x
\frac{}{}
\rc,
\no
B_{Af}^{gg} &=&
\,
\frac{32}{\e^2}
\lc
-  \pld
+ 2 \, x
-1
- \frac{11}{9}\, \dx
\rc
\no &&
+ \frac{8}{\e}
\lc
- \frac{8}{3}\, \lpld
+ \left( \frac{16}{3}\, x - \frac{8}{3} \right) \ln (1-x)
+ 3\, \pld
\right.
\no &&
\phantom{+\frac{8}{\e}}
\left.
+ \left( 6 \, x - \frac{8}{3\, (1-x)} - 2 \right) \ln x
- \frac{26}{3} \, x
+ \frac{20}{3}
+ \left( \frac{271}{27} \
         - \frac{8}{3}\, \zeta(2)
 \right) \dx
\rc ,
 \no
B_{Ff}^{gg} &=&
\,
\frac{8}{\e^2}
\lc
\frac{}{}
 4 \, (1+x) \ln x + 10 \, (1-x)
\frac{}{}
     \rc
\no &&
+\frac{4}{\e}
 \lc
\frac{}{}
8 \, (1+x) \, \li (1-x)
+8 \,   (1+x)\, \ln x \ln (1-x)
+ 6 \, (1+x)\, \ln^2x
\right.
\no &&
\phantom{+\frac{4}{\e}}
\left.
+ 20 \, (1-x) \ln (1-x)
+ (6 - 14 \,x) \ln x
+ 22 \, (x-1)
+  \dx
\frac{}{}
\rc,
\no
B_{AA}^{gg} &=&
\, \frac{4}{\e^2}
\lc
16\, \lpld
+ 16 \, (1-2\, x) \ln (1-x)
+ 22\, \pld
\right.
\no &&
\phantom{ \frac{8}{\e^2} }
\left.
- 8\, \left(3 + \frac{1}{1-x} \right) \ln x
+ 20 \, x - 42  + \dx \left(\frac{121}{9} - 8 \, \zeta(2) \right)
\rc
\no &&
+\frac{2}{\e}
\lc
-32\,(1+x) \li (1-x) +
\left( -16 \, x - \frac{8}{1+x} - 8 \right) \li (-x)
\right.
\no && \phantom{\frac{2}{\e}}
+ \left(-16 \, x + 4\, \pld - \frac{4}{1+x} \right) \zeta(2)
+ \left(-16 \, x - \frac{8}{1+x} - 8 \right) \ln x  \ln (1+x)
\no && \phantom{\frac{2}{\e}}
+24\, \llpld
+ \frac{88}{3}\, \lpld
+ \left( - 48\, x  + 24 \right) \ln^2 (1-x)
\no && \phantom{\frac{2}{\e}}
+\left( -48 \, x + \frac{8}{1-x} - 24 \right) \ln x \ln (1-x)
+\left(\frac{2}{1+x} - 32 - \frac{10}{1-x} \right) \ln^2 x
\no && \phantom{\frac{2}{\e}}
+ \left(\frac{208}{3} \, x - \frac{320}{3} \right) \ln (1-x)
+ \left(-14\, x + \frac{88}{3\,(1-x)} - 38 \right) \ln x
- 43 \pld
\no
&& \phantom{\frac{2}{\e}}
\left.
+ \frac{119}{3} - \frac{29}{3} \,x
+\dx \left( \frac{88}{3} \, \zeta(2) + 10 \, \zeta(3) - \frac{1663}{27}
\right)
\rc .
\nonumber
\end{eqnarray}
Here the function $\li (y)$ stands for the dilogarithm which can
be found in \cite{X28}.
After substitution of the one-loop order coefficients
$\beta_0$ (\ref{2.15}), $z_\alpha$ (\ref{smallzalpha}),
the Mellin transforms of
$P_{k,ij}^{(0)}$ $\left( = - \gamma_{k,ij}^{(0),m} \right)$
(\ref{3.4})--(\ref{3.7})
 and $a_{k,ij}^{(1)}$
(\ref{3.8})--(\ref{3.11})
into the algebraic expressions for
$\hat{A}_{k,ij}$ in
(\ref{2.21})--(\ref{2.27}) one can equate the latter with the
results obtained for $\hat{A}_{k,ij}$ as presented
above in (\ref{3.13})--(\ref{3.16}).
Notice that for $a_{k,qq}^{(1)} (k=NS,S)$ (\ref{3.8}) one has to choose
the `t Hooft Veltman 
prescription
which implies that we have to subtract from expression (\ref{3.8}) the 
constant $z_{qq} = - 8 C_F (1-x)$ (see below (\ref{3.12})).
{}From this one infers the two-loop contribution to the
anomalous dimensions
which are the unknown coefficients
in eqs.~(\ref{2.21})--(\ref{2.27}).
After performing the inverse Mellin transform we get
the splitting functions.
As has been already mentioned at the beginning of this section 
one has now to perform an additional renormalization to undo the 
unwanted effect of the $\gamma_5$-matrix prescription. This is achieved by multiplying
the OMEs $A_{k,qq}$ (k=NS,S) by an additional renormalization constant
$Z_{k,qq}=1 + \frac{g^2}{16 \pi^2}\, z_{aa}$(k=NS,S) with $z_{qq}$ given above.
This procedure will modify 
$P_{S,qg}^{(1)}$ and $P_{S,gq}^{(1)}$ so that one gets the splitting functions in the usual 
\MSbs.
The non-singlet splitting function $P_{NS,qq}^{(1)}$ is the
same as obtained in the spin-averaged case (see e.g.~\cite{X16,X17}).
In order to obtain the singlet splitting function
$P_{S,qq}^{(1)}$ one has to add to
$P_{NS,qq}^{(1)}$ the quantity $P_{PS,qq}^{(1)}$ given below.
\begin{eqnarray}
P_{PS,qq}^{(1)} &= &
C_F \,  T_f \,
 \left[
-16  (1 + x) \ln^2 x
-16  (1 - 3 x) \ln x
+ 16 (1 -  x)
\right].
\label{pqq1}
\end{eqnarray}
Furthermore we have the singlet splitting functions
\begin{eqnarray}
P_{S,qg}^{(1)} &= &
\phantom{+}
4 \, C_A \, T_f \;
 \left[
-8 (1 + 2 x) \li(-x)
- 8 \zeta(2)
-8 (1 + 2 x) \ln x \ln (1+x)
 \right.
\no
&&
\phantom{+}
\phantom{ 4 \, C_A \, T_f \;  }
+ 4 (1 - 2 x) \ln^2 (1-x)
-4 (1 + 2 x) \ln^2 x
\no
&&
\phantom{+}
\phantom{ 4 \, C_A \, T_f \;  }
\left.
-16 \,(1 - x) \ln (1-x)
+ 4 \, (1 + 8 x) \ln x
- 44 \, x + 48 \right]
     \no
     &&
+
  4 \, C_F \, T_f \,
\left[
8 \, ( 1 - 2 x ) \zeta(2)
-4 \, (1 - 2 x) \ln^2 (1-x)
\right.
\no
&&
\phantom{+}
\phantom{ 4 \, C_F \, T_f \;  }
+ 8 \, (1 - 2 x)
\ln x
\ln (1-x)
- 2\, (1 - 2 x) \ln^2 x
\no
&&
\phantom{+}
\phantom{ 4 \, C_F \, T_f \;  }
\left.
+ 16 \, (1 - x) \ln (1-x)
- 18 \, \ln x - 44 + 54 x \right],
\label{pqg1}
\end{eqnarray}
\begin{eqnarray}
P_{S,gq}^{(1)} &= &
\phantom{+}
C_A C_F \, \left[
16\, (2 + x) \li(-x)
+ 16\, x\, \zeta(2)
+8\, (2 - x) \ln^2 (1-x)
\right.
\no &&
\phantom{+}
\phantom{C_A C_F \,}
+ 16 \,(2 + x) \ln x \ln (1+x)
+ 8\, (2 + x) \ln^2 x
\no &&
\phantom{+}
\phantom{C_A C_F \,}
+ 16\, (x-2)
\ln x \ln (1-x)
+ \left( \frac{80}{3} + \frac{8}{3} x \right)
\ln (1-x)
\no &&
\phantom{+}
\phantom{C_A C_F \,}
\left.
+8 \,(4 - 13 \,x) \ln x
+\frac{328}{9} + \frac{280}{9} x
\right]
\no &&
+
  C_F^2 \;
\left[
 8\, (x -2) \ln^2 (1-x)
- 4\, (x - 2) \ln^2 x
- 68 + 32 x
\right.
\no && \phantom{C_F^2 \; [}
\left.
- 8 \, (x+2) \ln (1-x)
-4\, (4 -x) \ln x
\right]
\no &&
+
C_F T_f \,
 \left[
       -\frac{32}{9} (4+x)+\frac{32}{3} (x-2) \lmx
\right],
\label{pgq1}
\end{eqnarray}
\begin{eqnarray}
P_{S,gg}^{(1)} &=&
C_A^2 \,
\left[
\left(
64 x + \frac{32}{1 + x}  + 32
\right) \li(-x)
+
 \left(
        64 x - 16 \pld + \frac{16}{1 + x}
\right)
\zeta(2)
\right.
\no
&&
\phantom{ C_A^2 \, }
+\left(
 \frac{8}{1 - x} -  \frac{8}{1+x} + 32
\right)
\ln^2 x
+
\left( 64 x + \frac{32}{1 + x} + 32 \right) \ln x \ln (1+x)
\no
&&
\phantom{ C_A^2 \, }
+ \left( 64 x - \frac{32}{1 - x} -32 \right)
\ln x \ln (1-x)
+ \left(
\frac{232}{3}
- \frac{536}{3} x
 \right) \ln x
\no
&&
\left.
\phantom{ C_A^2 \, }
+ \frac{536}{9} \left( \frac{1}{1 - x} \right)_+
- \frac{388}{9} x
-\frac{148}{9}
+
\delta(1-x)
( 24 \zeta(3) + \frac{64}{3} )
\right]
\no
&&
+
C_A T_f \,
\left[
- \frac{160}{9} \left( \frac{1}{1 - x} \right)_+
-\frac{32}{3} \left( 1 + x \right) \ln x
-\frac{448}{9}
+ \frac{608}{9} x
\right.
\no
&&
\phantom{C_A T_f \,\,}
\left.
- \frac{32}{3} \delta(1-x)
\right]
\no
&&
+C_F T_f \,
\left[
- 16(1 + x) \ln^2 x
+ 16 (x - 5) \ln x
- 80 (1 - x)
\right.
\no
&&
\phantom{+C_A T_f \,}
\left.
- 8 \delta(1-x)
{}^{}
\right].
\label{pgg1}
\end{eqnarray}
For practical purposes, and the discussion of
the results obtained above in the next section,
it is also useful to present the one- and two-loop
anomalous dimensions which are related to the splitting functions
via the Mellin transform
\begin{equation}
\gamma_{k,ij}^{m} = - \int_0^1 dx \, x^{m-1} \,
P_{k,ij}(x).
\label{3.21}
\end{equation}
The one-loop contribution to the anomalous dimensions become
\begin{eqnarray}
\gamma_{NS,qq}^{(0), m} &=&
\gamma_{S,qq}^{(0), m}  =
C_F \, \left[8 S_1(m-1) + \frac{4}{m} + \frac{4}{m+1} - 6  \right],
\label{3.22}
\\
\gamma_{S,qg}^{(0), m} &=&
T_f \,
\left[\frac{8}{m} - \frac{16}{m+1} \right],
\label{3.23}
\\
\gamma_{S,gq}^{(0), m} &=&
C_F \, \left[\frac{4}{m+1} - \frac{8}{m} \right],
\label{3.24}
\\
\gamma_{S,gg}^{(0), m} &=&
C_A \left[8 S_1(m-1) - \frac{8}{m} + \frac{16}{m+1} - \frac{22}{3} \right]
+\frac{8}{3} T_f .
\label{agg0}
\end{eqnarray}
The two-loop non-singlet anomalous dimension
$\gamma_{NS,qq}^{(1),m}$  is the
same as found for the spin averaged
operator (see e.g.~ \cite{X15,X16}).
To obtain  the singlet anomalous dimension
$\gamma_{S,qq}^{(1),m}$ one has to add $\gamma_{NS,qq}^{(1),m}$
the quantity $\gamma_{PS,qq}^{(1),m}$
which reads
\begin{eqnarray}
\gamma_{PS,qq}^{(1),m}
&=&
C_F T_f \, 16 \,
\left[
 \frac{2}{(m+1)^3}
+ \frac{3}{(m+1)^2}
+ \frac{1}{(m+1)}
+ \frac{2}{m^3}
- \frac{1}{m^2}
- \frac{1}{m}
\right].
\label{aqq1}
\end{eqnarray}
Furthermore we have the  singlet anomalous dimensions
\begin{eqnarray}
\gamma_{S,qg}^{(1),m}
&=&
16 \,
C_A T_f \;
\left[
-  \frac{S_1^2(m-1)}{m}
+  \frac{2\, S_1^2(m-1)}{m+1}
-  \frac{2\, S_1(m-1)}{m^2}
+ \frac{4 \, S_1(m-1)}{(m+1)^2}
\right.
\no
&&
\phantom{ C_A T_f \; 16 \, }
- \frac{ S_2(m-1)}{m}
+ \frac{2\,  S_2(m-1)}{m+1}
- \frac{2 \, \tilde{S}_2(m-1)}{m}
+ \frac{4 \, \tilde{S}_2(m-1)}{m+1}
\no
&&
\left.
\phantom{ C_A T_f \; 16 \, }
- \frac{4}{m}
+ \frac{3}{m+1}
- \frac{3}{m^2}
+ \frac{8}{(m+1)^2}
+ \frac{2}{m^3}
+\frac{12}{(m+1)^3}
\right]
\no
&&
+  \,
8 \, C_F T_f \;
\left[
\frac{2 \, S_1^2(m-1)}{m}
- \frac{4 \, S_1^2(m-1)}{m+1}
- \frac{2 \, S_2 (m-1)}{m}
+\frac{4 \, S_2(m-1)}{m+1}
\right.
\no
&&
\phantom{+ \, C_F T_f \;8 \, \;}
\left.
+ \frac{14}{m}
- \frac{19}{m+1}
- \frac{1}{m^2}
- \frac{8}{(m+1)^2}
- \frac{2}{m^3}
+ \frac{4}{(m+1)^3}
\right],
\label{aqg1}
\end{eqnarray}
\begin{eqnarray}
\gamma_{S,gq}^{(1),m}
&=&
8 \, C_A C_F \;
\left[
-\frac{2\, S_1^2(m-1)}{m}
+\frac{S_1^2(m-1)}{m+1}
+\frac{16\, S_1(m-1)}{3 m}
- \frac{5 S_1(m-1)}{3( m+1)}
\right.
\no && \phantom{ C_A C_F \,\, }
+ \frac{2 S_2(m-1)}{m}
-\frac{S_2(m-1)}{m+1}
+\frac{4\tilde{S}_2(m-1)}{m}
-\frac{2\tilde{S}_2(m-1)}{m+1}
-\frac{56}{9 m}
\no && \phantom{ C_A C_F \,\, }
\left.
-\frac{20}{9 (m+1)}
+\frac{28}{3 m^2}
-\frac{38}{3 (m+1)^2}
-\frac{4}{m^3}
- \frac{6}{(m+1)^3}
\right]
\no
&&
+ 4 \, C_F^2 \;
\left[
 \frac{4 \, S_1^2(m-1)}{m}
-\frac{2 \, S_1^2(m-1)}{m+1}
-\frac{8 \, S_1(m-1)}{m}
+\frac{2 \, S_1(m-1)}{m+1}
\right.
\no && \phantom{+ \, C_F^2 \; 4 \, }
+\frac{8 \, S_1(m-1)}{m^2}
-\frac{4 \, S_1(m-1)}{(m+1)^2}
+\frac{4 \, S_2(m-1)}{m}
-\frac{2 \, S_2(m-1)}{m+1}
\no && \phantom{+ \, C_F^2 \; 4 \,}
\left.
+\frac{15}{m}
-\frac{6}{m+1}
-\frac{12}{m^2}
+ \frac{3}{(m+1)^2}
+ \frac{4}{m^3}
- \frac{2}{(m+1)^3}
\right]
\no
&&
+ 32 \, C_F T_f \;
\left[
- \frac{2 \, S_1(m-1)}{3 \, m}
+ \frac{S_1(m-1)}{3 \, (m+1)}
+ \frac{7}{9 \, m}
\right.
\no && \phantom{+ C_F T_f \;32  \,}
\left.
- \frac{2}{9 \, (m+1)}
- \frac{2}{3 \, m^2}
+ \frac{1}{3 \, (m+1)^2}
\right],
\label{agq1}
\end{eqnarray}
\begin{eqnarray}
\gamma_{S,gg}^{(1),m}
&=&
4 \, C_A^2
\left[
\frac{134}{9} \, S_1(m-1)
+ \frac{8  \, S_1(m-1)}{m^2}
- \frac{16  \, S_1(m-1)}{(m+1)^2}
\right.
\no && \phantom{C_A^2 \, 4 \,}
+ \frac{8  \, S_2(m-1)}{m}
- \frac{16  \, S_2(m-1)}{m+1}
+ 4 \, S_3(m-1)
\no && \phantom{C_A^2 \, \,}
- 8  \, S_{1,2}(m-1)
- 8  \, S_{2,1}(m-1)
+ \frac{8  \,\tilde{S}_2(m-1)}{m}
- \frac{16 \,\tilde{S}_2(m-1)}{m+1}
\no && \phantom{C_A^2 \, \,}
+ 4 \, \tilde{S}_3(m-1)
- 8  \, \tilde{S}_{1,2}(m-1)
- \frac{107}{9 \, m}
+ \frac{241}{9 \, (m+1)}
\no && \phantom{C_A^2 \, \,}
\left.
+ \frac{58}{3 \, m^2}
- \frac{86}{3 \, (m+1)^2}
- \frac{8}{m^3}
- \frac{48}{(m+1)^3}
- \frac{16}{3}
\right]
\no
&&
+ 32 \, C_A T_f \,
\left[
\frac{-5 \, S_1(m-1)}{9} + \frac{14}{9 m} -
\frac{19}{9 \, (m+1)} - \frac{1}{3 \, m^2} -
\frac{1}{3 \, (m+1)^2} + \frac{1}{3}
\right]
\no
&& +8 \, C_F T_f \,
\left[
-\frac{10}{m+1} + \frac{2}{(m+1)^2} + \frac{4}{(m+1)^3}
+ 1 + \frac{10}{m} - \frac{10}{m^2} + \frac{4}{m^3}
\right],
\label{agg1}
\end{eqnarray}
where we have introduced the following notations\footnote{Notice that only for $m$ odd the above anomalous dimensions correspond to physical operators and that we have implicitly multiplied them by the factor $(1-(-1)^m)/2$}
\begin{eqnarray}
S_k (m-1) &=& \sum_{i=1}^{m-1} \frac{1}{i^k},
\no
\tilde{S}_k (m-1) &=& \sum_{i=1}^{m-1} \frac{(-1)^i}{i^k},
\no
S_{k,l}(m-1) &=&  \sum_{i=1}^{m-1} \frac{1}{i^k} \, S_l(i),
\no
\tilde{S}_{k,l}(m-1) &=&  \sum_{i=1}^{m-1} \frac{1}{i^k} \,
\tilde{S}_l(i) .
\nonumber
\end{eqnarray}
To check our results for the two-loop splitting functions
(anomalous dimensions) we have also used the BPHZ method
\cite{X19}
as mentioned at the end of section 2. Here we renormalized
the OMEs graph by graph and found finally the same
results as listed in (\ref{pqq1})--(\ref{pgq1}).
As already mentioned in the beginning the above
splitting functions and anomalous dimensions have been
calculated in the \MSbs.
If one prefers another scheme the corresponding anomalous
dimensions are related to the  \MSb\ ones in the following way
\begin{eqnarray}
\gamma_{NS,qq} &=& \bar{\gamma}_{NS,qq}
+ \beta(g) Z_{NS} \frac{d \, Z_{NS}^{-1}}{d g},
\label{3.30}
\\
\gamma _{S,ij} &=& Z_{il} \bar{\gamma}_{lm}
\left( Z^{-1}\right)_{mj} + \beta(g) Z_{il}
\frac{d \, (Z^{-1})_{lj}}{d g},
\label{3.31}
\end{eqnarray}
where $\bar{\gamma}_{k,qq}$ $(k=NS,S)$ denotes the anomalous
dimension in the \MSbs\ and $Z_{NS}$, $Z_{ij}$ are finite
operator renormalization constants. Up to order $g^2$
they can be expressed as follows
\begin{eqnarray}
Z_{NS} &=& 1 + \frac{g^2}{16 \pi^2} \, z_{qq},
\label{3.32}
\\
Z &=& \left(
\begin{array}{cc}
1 + \frac{g^2}{16 \pi^2} z_{qq} & \frac{g^2}{16 \pi^2} z_{qg} \\
\frac{g^2}{16 \pi^2} z_{gq}     & 1 + \frac{g^2}{16 \pi^2} z_{gg}
\end{array}
\right).
\label{3.33}
\end{eqnarray}
Substitution of eqs.~(\ref{3.32}), (\ref{3.33}) into
eqs.~(\ref{3.30}), (\ref{3.31}) yields
\begin{eqnarray}
\gamma_{NS,qq}^{(1)} &=& \bar{\gamma}_{NS,qq}^{(1)} +
 2 \beta_0 z_{qq},
\label{gansqq}
\\
\gamma_{S,qq}^{(1)} &=& \bar{\gamma}_{S,qq}^{(1)} +
2 \beta_0 z_{qq} + z_{qg} \bar{\gamma}_{S,gq}^{(0)} -
 \bar{\gamma}_{S,qg}^{(0)} z_{gq},
\label{3.35}
\\
\gamma_{S,qg}^{(1)} &=&
\bar{\gamma}_{S,qg}^{(1)} +
2 \beta_0 Z_{qg} + z_{qg}
 \left(  \bar{\gamma}_{S,gg}^{(0)} -
         \bar{\gamma}_{S,qq}^{(0)}
\right)
+  \bar{\gamma}_{S,qg}^{(0)}
\left( z_{qq}  - z_{gg} \right),
\label{3.36}
\\
\gamma_{S,gq}^{(1)} &=&
\bar{\gamma}_{S,gq}^{(1)} +
2 \beta_0 z_{gq}
+ z_{gq}
 \left(  \bar{\gamma}_{S,qq}^{(0)} -
         \bar{\gamma}_{S,gg}^{(0)}
\right)
+  \bar{\gamma}_{S,gq}^{(0)}
\left( z_{gg}  - z_{qq} \right),
\label{3.37}
\\
\gamma_{S,gg}^{(1)} &=&
\bar{\gamma}_{S,gg}^{(1)} +
2 \beta_0 z_{gg} + z_{gq} \bar{\gamma}_{S,qg}^{(0)} -
 \bar{\gamma}_{S,gq}^{(0)} z_{qg} .
\label{gansgg}
\end{eqnarray}
Before finishing this section we want to make a comment
on the spin splitting functions and the
anomalous dimensions calculated
above. Two of them, i.e.,
$P_{PS,qq}^{(1)}$ $(\gamma_{PS,qq}^{(1)})$
and  $P_{S,qg}^{(1)}$ $(\gamma_{S,qg}^{(1)})$
have been already calculated in the literature
\cite{X9}.
They were obtained via mass factorization of the partonic
cross sectoin of the subprocesses $\gamma^* + q \rightarrow
q + q + \bar{q}$ and  $\gamma^* + g \rightarrow
g + q + \bar{q}$ including the virtual corrections to
$\gamma^* + g \rightarrow q + \bar{q}$.
The result for $P_{PS,qq}^{(1)}$
(\ref{pqq1}) agrees with eq.~(3.37) in \cite{X9}.
However, the expression
for $P_{S,qg}^{(1)}$
in (\ref{pqg1})  differs from the
one obtained in eq.~(3.38) of \cite{X9} by a finite
renormalization, i.e.,
\begin{equation}
P_{S,qg}^{(1)}(\cite{X9}) -
P_{S,qg}^{(1)}(\ref{pqg1})  =
P_{S,qg}^{(0)} \otimes z_{qq},
\label{3.39}
\end{equation}
with
\begin{equation}
z_{qq} = -8 \, C_F \, (1-x) ,
\label{3.40}
\end{equation}
and $\otimes$ denotes the convolution symbol
\begin{equation}
(f \,  \otimes \,  g)(x) = \int_0^1 dx_1 \, \int_0^1 dx_2 \,
\delta( x- x_1 x_2) \, f(x_1) \, g(x_2) .
\end{equation}
 This difference can be attributed to the fact that in \cite{X9} one has forgotten to perform
 the additional renormalization given by $Z_{k,qq}$ above eq. \ref{pqq1} due to the `t Hooft Veltman prescription \cite{X21}.
 If one performs the same renormalization to the transition functions 
 $\Gamma_{qq}^{NS}$ (3.46) and $\Gamma_{qq}^{S}$ (3.47) in \cite{X9} one obtains
 the same splitting function $P_{S,qg}^{(1)}$ as presented in \ref{pqq1}. 
 Notice that this finite renormalization also affects the second term in the coefficient function 
 $\overline{C}_g$ in eq. (A.5) of \cite{X9}.

The above splitting functions, which are calculated in the
\MSbs, have to be combined with the
quark and gluon
coefficient functions (\ref{2.10})
computed in the same scheme in order to perform a complete
next-to-leading order analysis. The quark
coefficient function can be found in
\cite{X9, X11}
and it equals to
\begin{eqnarray}
\eco_q(x,Q^2,\mu^2) &=&
\dx
+ \frac{g^2}{16 \pi^2} C_F \,
\left[ \lc 4 \pld - 2 - 2 \, x
\right.
\right.
\no &&
\left.
+ 3 \, \dx \rc \ln \frac{Q^2}{\mu^2}
+ 4 \, \lpld
- 2 \, ( 1+x) \lmx
\no &&
\left.
- 2 \, \frac{1+x^2}{1-x} \ln x
- 3 \pld + 4 + 2 \, x - \dx (9+4 \zeta (2) )
\right].
\end{eqnarray}
The gluon coefficient function (see e.g.~\cite{X9,X29})
gets the form
\begin{eqnarray}
\tilde{E}_g(x,Q^2,\mu^2) &=&
\frac{g^2}{16 \pi^2} T_f \left[
(8 \, x - 4 ) \ln \frac{Q^2}{\mu^2}
+ (8 \, x - 4) \ln (1-x)
\right.
\no
&& \phantom{\frac{g^2}{16 \pi^2} T_f \; }
\left.
- (8 \, x - 4) \ln x + 12 - 16 \, x
\right].
\end{eqnarray}
The Mellin transforms of $\eco_k \; (k=q,g)$ become
\begin{eqnarray}
\eco_q^m(Q^2,\mu^2) &=&
1 + \frac{g^2}{16 \pi^2} C_F \,
\left[ \left(3 - \frac{2}{m} - \frac{2}{m+1} - 4 S_1(m-1) \right)
\ln \frac{Q^2}{\mu^2}
\right.
\no &&
\left.
+ \left( \frac{2}{m} + \frac{2}{m+1} + 3 \right) S_1(m-1)
+ 4 \, S_{1,1}(m-1)
\right.
\no &&
\left.
- 4\, S_2(m-1) + \frac{6}{m} - 9
\right],
\\
\eco_g^m (Q^2, \mu^2) &=&
\frac{g^2}{16 \pi^2} T_f \left[
\left( \frac{8}{m+1} - \frac{4}{m} \right) \ln \frac{Q^2}{\mu^2}
+ \left( \frac{4}{m} - \frac{8}{m+1} \right) S_1 (m-1)
\right.
\no && \left.
+ \frac{4}{m} - \frac{8}{m+1} \right].
\end{eqnarray}
Notice that the first moment $\eco_q^1 = 1 - 3 (g^2/(16 \pi^2)) C_F$
agrees with eq.~6 of \cite{X10}.
Furthermore we have $\eco_g^1 = 0$ (see \cite{X11}).
Both properties are characteristic of our choice of the
$\gamma_5$-prescription and the fact  that the
anomalous dimensions are calculated in the
\MSbs.

\section{Properties of the spin anomalous dimensions}
In this section we will discuss some of the properties of the splitting
functions and anomalous dimensions which have been calculated in the
last section. Let us start with the first moments of the
spin anomalous dimensions in the \MSbs.
\begin{eqnarray}
\gamma _{NS,qq}^{(0),1} &=& 0  \qquad \qquad  \,
\gamma _{NS,qq}^{(1),1} \;\;=\;\; 0,
\label {4.1}
\\
\gamma _{S,qq}^{(0),1} &=& 0  \qquad \qquad \;
\gamma _{S,qq}^{(1),1} \;\;=\;\;  24 \, C_F \, T_f,
\\
\label {4.2}
\gamma _{S,qg}^{(0),1} &=& 0 \qquad \qquad \; \; \;
\gamma _{S,qg}^{(1),1} \;\;=\;\; 0,
\\
\label {4.3}
\gamma _{S,gq}^{(0),1} & =& -6 \, C_F  \qquad  \; \;
\gamma _{S,gq}^{(1),1} \;\;=\;\; 18 \, C_F^2
- \frac{142}{3} \, C_A \, C_F + \frac{8}{3} \, C_F \, T_f,
\label {4.4}
\\
\gamma _{S,gg}^{(0),1} & =& -2 \, \beta_0 \;\; = \;\;
- \left( \frac{22}{3} \, C_A - \frac{8}{3} \, T_f \right),
\\
\gamma _{S,gg}^{(1),1} & =& -2 \, \beta_1 \;\; =\;  \;
- \frac{68}{3} \,  C_A^2 + 8 \, C_F \, T_f  +
\frac{40}{3} \, C_A \, T_f,
\end{eqnarray}
where $\beta_0$ and $\beta_1$ are the first and second
order coefficients in the perturbation series of the
$\beta $-function (\ref{2.15}).

In the above we have assumed that there is one flavour only
in the fermion loops of the OME graphs.
If there are more light flavours the $T_f$ in the above expressions
have to be multiplied by the number of light flavours indicated by
$n_f$ (see (\ref{2.16}), (\ref{2.17})).
The vanishing of the first moment of the non-singlet
anomalous dimension follows from the conservation of the
axial vector current $R_{NS,qq}^\mu (x)$. The value of the
singlet anomalous dimension $\gamma_{S,qq}^{(1),1}$ was already
calculated in \cite{X29}.
It is due to the anomaly
of the singlet axial vector current
$R_{S,q}^\mu (x)$ which contributes via the
triangular fermion loops to  $\gamma_{S,qq}^{(1),1}$ in second
order perturbation theory.
The vanishing of $\gamma_{S,gq}^{(1),1}$
was shown on general grounds in \cite{X30},
see also \cite{X31}.
{}From the last reference we also infer (see eq.~(22)
in \cite{X31}) that  $\gamma_{S,gg}^{(1),1} = - 2 \beta_1$ ,
provided the anomalous dimension is calculated in the
\MSbs. Finally we want to investigate an
interesting relation which is conjectured
for an ${\mathcal{N}} = 1$ supersymmetric Yang-Mills field theory.
It can be derived from QCD by putting the colour factors
$C_F = C_A = N$ and $T_f = N/2$ \cite{X32}.
The relation reads as follows. First define
\begin{equation}
\delta   \gamma = \gamma_{S,qq} + \gamma_{S,gq} -  \gamma_{S,qg}-
 \gamma_{S,gg}.
\label{4.6}
\end{equation}
For an ${\mathcal{N}}=1$ supersymmetric Yang-Mills field theory one has
\begin{equation}
\delta  \gamma = 0,
\label{4.7}
\end{equation}
provided $\delta  \gamma $ is calculated in a renormalization
scheme which preserves the supersymmetric Ward identities.
In many cases one has shown that
at least up to two loops $n$-dimensional
reduction  is a regularization method which respects
the supersymmetric Ward identities.
Therefore a renormalization scheme where the pole terms plus the
additional constants $\gamma_E$ (Euler constant) and $\ln 4 \pi$
are subtracted (\MSbs) will respect these Ward identities
too. In lowest order,
 where there is no difference between
$n$-dimensional regularization and $n$-dimensional reduction,
the above relation holds for the spin as well as spin averaged
anomalous dimensions.
If one assumes that the two-loop anomalous dimensions
calculated in the
two regularization schemes ($n$-dimensional reduction and
$n$-dimensional regularization) are related to
each other via a finite renormalization
one can derive the following relation \cite{X33}
\begin{equation}
\delta \,
\gamma_{\rm RED}^{(1),m} -
\delta \,
\gamma_{\rm REG}^{(1),m}  =
\left(
2 \, \beta_0 - \gamma_{S,qg}^{(0),m} -
\gamma_{S,gq}^{(0),m}
\right) \left(\delta \, a_{\rm RED}^{(1),m}  -
              \delta \, a_{\rm REG}^{(1),m}
       \right),
\label{4.8}
\end{equation}
with
\begin{equation}
\delta \, a^{(1)}=
a_{S,qq}^{(1)} + a_{S,gq}^{(1)} - a_{S,qg}^{(1)} - a_{S,gg}^{(1)},
\label{4.9}
\end{equation}
where the terms $a^{(1)}_{S,ij, {\rm REG}}$ and
$a^{(1)}_{S,ij, {\rm RED}}$ are the non-pole parts of the OMEs in
(\ref{2.21})--(\ref{2.27}) which are calculated using
$n$-dimensional regularization and
$n$-dimensional reduction respectively.
Equation (\ref{4.8}) can be easily derived
from eqs.~(\ref{gansqq})--(\ref{gansgg}) by putting
$\delta \gamma^{(0)} = 0$ and
$z_{ij}  = a_{S,ij,{\rm REG}}$--$a_{S,ij,{\rm RED}}$.
If one makes the additional assumption
$\delta \gamma^{(1)} =
\delta \gamma^{(1)}_{\rm RED} = 0$ (\ref{4.7}) relation
(\ref{4.8}) turns out to be valid for the
two-loop spin averaged anomalous dimensions which
is checked in \cite{X33}.

\noindent
In the case of the spin anomalous dimensions we obtain
the following results
\begin{eqnarray}
a_{S,qq,{\rm REG}}^{(1)} - a_{S,qq,{\rm RED}}^{(1)} &=& N
\left[-2 + 2\, x + \dx \right],
\label{4.10}
\\
a_{S,qg,{\rm REG}}^{(1)} - a_{S,qg,{\rm RED}}^{(1)} &=& 0,
\label{4.11}
\\
a_{S,gq,{\rm REG}}^{(1)} - a_{S,gq,{\rm RED}}^{(1)} &=& 0,
\label{4.12}
\\
a_{S,gg,{\rm REG}}^{(1)} - a_{S,gg,{\rm RED}}^{(1)} &=&
N \left[ \frac{1}{3} \, \dx \right].
\label{4.13}
\end{eqnarray}
If we assume that $\delta \, \gamma _{\rm RED}^{(1)} = 0 $ then
(from (\ref{4.8}))
\begin{equation}
 \delta \, \gamma_{\rm REG}^{(1),m}=
N^2 \left (4 - \frac{8}{m^2}
  + \frac{8}{(m+1)^2}
  - \frac{28}{3} \frac{1}{m}
  + \frac{44}{3} \frac{1}{m+1} \right ),
\label{4.14}
\end{equation}
 which is in agreement with the result of our calculation
 derived from eqs.~(\ref{aqq1})-(\ref{agg1}).
 To understand why the supersymmetric relation (\ref{4.7})
 is valid 
 we investigated the findings in \cite{X33} for the spin averaged anomalous dimensions
 and found a surprising result. 

Since all external legs of the
OMEs are put off shell one can split
$\hat{G}_{k,iq}^{m}$ (\ref{3.1}) and
$\hat{G}_{k,ig}^{m,\mu, \nu}$ (\ref{3.2}) into a
so-called physical and unphysical part. In the case of
$\hat{G}_{k,iq}^m$ the former part is proportional to $\Ds$ whereas the
latter part is multiplied by $\ps$.
This property holds for the spin as well
as spin averaged operators.
In the spin case $\hat{G}_{S,ig}^{m,\mu \nu}$ has
a physical part only which is proportional to $\e^{\mu \nu \alpha \beta}
\Delta_\alpha p_\beta$.
However, for the spin averaged case one also encounters unphysical
parts in the OMEs.
Here the physical part is
the coefficient of the tensor
$g^{\mu \nu}- (\Delta_\mu p_\nu + p_\mu \Delta_\nu)/(\Delta p) +
\Delta_\mu \Delta_\nu p^2/ (\Delta p)^2$.
Limiting ourselves to the physical parts of the
non-pole terms $a_{S,ij}^{(1)}$ we find the following results.
The spin averaged OMEs satisfy the relation (Feynman gauge)
\footnote{Notice that the gauge dependent terms cancel in
$\delta a_{\rm RED}^{(1),m} - \delta a_{\rm REG}^{(1),m}$.
}
\begin{equation}
\delta a_{\rm RED}^{(1)} = 0,
\label{4.16}
\end{equation}
whereas the spin OMEs lead to
\begin{equation}
\delta a_{\rm RED}^{(1)} = N \left[ - 4 + 4\, x \right].
\label{4.17}
\end{equation}
which equals the ABJ-anomaly.
Property (\ref{4.16}) was not mentioned in \cite{X33}.
However, it might explain why $\delta \gamma_{\rm RED}^{(1)} = 0$ for the
spin averaged case since the physical part of the unrenormalized one-loop
OMEs already satisfy the supersymmetric relation. 
In the case of the spin anomalous dimensions the validity
of the supersymmetric relation (\ref{4.7}) can be attributed to the fact that 
the {\it ABJ}-anomaly only gets finite renormalizations due to the
choice of our $g^5$-matrix prescription \cite{X23}.
Since this anomaly is already of order $\alpha_s$ it implies that 
the ultraviolet divergencies in the unrenomralized OMEs will not be
affected up to order $\alpha_s^2$ so that the supersymmetric 
relation is still valid to two-loop order in perturbation theory.

\appendixon
\namedappendix{The operator vertices}
\noindent
In this appendix we present the twist-2 operator vertices.
All momenta are flowing into the operator vertex.

\subsection{Quark-(gluon) operator vertices}
The quark-antiquark  vertex is equal to
\begin{eqnarray}
O(p) &=& - \Ds \gamma_5 (\Delta p)^{m-1},
\label{lq2}
\end{eqnarray}
where $p$ denotes the momentum of the incoming quark line.

\noindent
The quark-quark-gluon vertex is given by
\begin{eqnarray}
O^{\mu}_a (p, q)  &=& - g \, T_a \, \Delta^{\mu} \Ds \gamma_5
 \sum_{i=0}^{m-2}(\Delta p)^{m-i-2} (-\Delta q)^i,
\end{eqnarray}
where $p$ and $q$ are the momenta of the incoming quark
and antiquark respectively.

\noindent
The quark-quark-gluon-gluon vertex equals
\begin{eqnarray}
O_{ab}^{\mu \nu}(p,q,r,s) &=&
  g^2 \, \Delta^\mu  \, \Delta^\nu
\Ds \gamma_5
\no
&&\times
\left[
T_a T_b \,
\left( (-1)^m  \sum_{j=0}^{m-3} \sum_{i=0}^j \,
(-1)^j  \ddp^i \ddq^{m-j-3} (\ddp + \dds)^{j-i}
\right)
\right.
\no
&&
\left.
-
T_b T_a \,
\left( \sum_{j=0}^{m-3} \sum_{i=0}^j \,
(-1)^j  \ddp^{m-j-3} \ddq^{j} (\ddq + \dds)^{j-i}
\right)
\right].
\end{eqnarray}

\subsection{Gluon-operator vertices}
The 2-gluon vertex is given by
\begin{equation}
O^{\mu \nu}_{a b}(p) =
 i \e^{\mu \nu \Delta p} (1 - (-1)^m) (\Delta p)^{m-1} \delta_{ab}.
\label{Atwogluon}
\end{equation}
The 3-gluon vertex is  equal to
\begin{eqnarray}
O_{abc}^{\mu \nu \rho}(p,q,r) &=&
g (1 - (-1)^m) \, f_{abc} \,
O^{\mu \nu \rho} (p,q,r),
\label{Athreegluon}
\end{eqnarray}
with
\begin{eqnarray}
O^{\mu \nu \rho} (p,q,r)
& = &
- (\e^{\mu \rho p \Delta} \Delta^{\nu} -
   \e^{\mu \nu p \Delta} \Delta^\rho
  ) (\Delta p)^{m-2}
- (\e^{\nu \rho \Delta q} \Delta^{\mu} +
   \e^{\mu \nu \Delta q} \Delta^\rho
  ) (\Delta q)^{m-2}
\no
&&
- (\e^{\nu \rho \Delta r} \Delta^{\mu} -
   \e^{\mu \rho \Delta r} \Delta^\nu
  ) (\Delta r)^{m-2}
\no
&&
+
\sumi (-\Delta p)^i (\Delta q)^{m-i-3} \Delta^\rho
(\e^{\nu p \Delta q} \Delta^\mu + \e^{\mu \nu \Delta q} (\Delta p))
\no
&&
-
\sumi (-\Delta r)^i (\Delta p)^{m-i-3} \Delta^\nu
(\e^{\mu p \Delta r} \Delta^\rho + \e^{\mu \rho \Delta p} (\Delta r))
\no
&&
-
\sumi (-\Delta r)^i (\Delta q)^{m-i-3} \Delta^\mu
(\e^{\nu \Delta q r} \Delta^\rho - \e^{\nu \rho \Delta q} (\Delta r)
).
\end{eqnarray}

The 4-gluon vertex equals
\begin{eqnarray}
O_{abcd}^{\mu \nu \rho \sigma}(p,q,r,s) &=&
i \, g^2 \,  (1- (-1)^m) \left[
f_{a b e} f_{c d e} O^{\mu \nu \rho \sigma}(p,q,r,s)
\right.
\no
&&
\left.
+ f_{a c e} f_{b d e} O^{\mu \rho \nu \sigma}(p,r,q,s)
- f_{a d e} f_{b c e} O^{ \rho \nu \mu \sigma}(r,q,p,s)
\right],
\end{eqnarray}
where
\begin{eqnarray}
&&
O^{\mu \nu \rho \sigma}(p,q,r,s) \; = \;
(\e^{\Delta \nu \rho \sigma} \Delta^\mu -
 \e^{\Delta \mu \rho \sigma} \Delta^\nu )
( \ddr + \dds)^{m-2}
\no &&
-
\Delta^\rho
(\e^{\nu \sigma \Delta s} \Delta^\mu
 -\e^{\mu \sigma \Delta s} \Delta^\nu
)
\sum_{i=0}^{m-3}(\ddr + \dds)^i
\dds^{m-i-3}
\no &&
+
\Delta^\sigma
(\e^{\rho \nu r \Delta } \Delta^\mu -
 \e^{\rho \mu r \Delta } \Delta^\nu
)
\sum_{i=0}^{m-3}(\ddp + \ddq)^{m-i-3} (-\ddr)^i
\no &&
+
\Delta^\nu
\sum_{i=0}^{m-3}
(\ddr + \dds)^{m-i-3}
(-\ddp)^i
(\e^{\mu \sigma \Delta p } \Delta^\rho -
 \e^{\mu \rho \Delta p} \Delta^\sigma
)
\no &&
+
\Delta^\mu
\sum_{i=0}^{m-3}
(\ddr + \dds)^{m-i-3} (-\ddq)^i
(\e^{\nu \sigma \Delta q} \Delta^\rho -
 \e^{\nu \rho \Delta q} \Delta^\sigma
)
\no &&
+
\Delta^\nu \Delta^\rho
\sum_{j=0}^{m-4}
\sum_{i=0}^j
\ddp^{m-j-4} (\ddp + \ddq)^{j-i} (-\dds)^i
(\e^{\Delta \sigma p s} \Delta^\mu +
  \e^{\mu \sigma \Delta s} \ddp
)
\no &&
-
\Delta^\mu \Delta^\rho
\sum_{j=0}^{m-4}
\sum_{i=0}^j
\ddq^{m-j-4} (\ddp + \ddq)^{j-i} (-\dds)^i
(\e^{\Delta \sigma q s} \Delta^\nu +
  \e^{\nu \sigma \Delta s} \ddq
)
\no &&
-
\Delta^\nu \Delta^\sigma
\sum_{j=0}^{m-4}
\sum_{i=0}^j
\ddp^{m-j-4}
(-\ddr)^i
(\ddp + \ddq)^{j-i}
(\e^{\Delta \mu p r} \Delta^\rho +
  \e^{\mu \rho \Delta p} \ddr
)
\no &&
+
\Delta^\mu \Delta^\sigma
\sum_{j=0}^{m-4}
\sum_{i=0}^j
\ddq^{m-j-4}
(-\ddr)^i
(\ddp + \ddq)^{j-i}
(\e^{\Delta \nu q r} \Delta^\rho +
  \e^{\nu \rho \Delta q} \ddr
).
\label{Afourgluon}
\end{eqnarray}

\namedappendix{The tensorial reduction}

\noindent
In this appendix we present a more detailed explanation of the
tensorial reduction of the tensor Feynman integrals into scalar integrals.

\noindent
According to the reading point method \cite{X20} we can put the
$\gamma^5$-matrix at the right end of the traces.
Then one can perform all straightforward simplifications of
the $\gamma$-matrix algebra inside the traces. Furthermore
we leave the $\gamma^5$-matrix untouched except that at the end
we take $Tr\left( \gamma^\alpha \gamma^\beta \gamma^\sigma
\gamma^\delta \gamma^5 \right) =
-4 \, i \, \epsilon ^{\alpha \beta \sigma \delta}$
In the case of the one-loop integrals the tensorial reduction
can be very easily achieved via the standard Feynman parameter
techniques. Since we do not need tensor integrals beyond rank two
it is sufficient to list the following integrals
\begin{eqnarray}
I_{ij} &=&
\int \frac{d^n q}{(2 \pi)^n} \,
\frac{( \Delta q)^m}{ \left[q^2\right]^i \left[(q-p)^2 \right]^j }
\no
&=& i \, S_n \,
\frac{ (-p^2)^{n/2}}{(p^2)^{i+j}}
\, ( \Delta p )^m \,
\frac{\Gamma \left(i+j-\frac{n}{2} \right)}{\Gamma (i) \Gamma (j)}
\; \int_0^1 d x \; x^m \; x^{n/2 - 1 -i} (1-x)^{n/2-1-j},
\label{B.1}
\\
I_{ij}^\mu &=&
\int \frac{d^n q}{(2 \pi)^n} \,
\frac{q^\mu \, ( \Delta q)^m}{ \left[q^2\right]^i \left[(q-p)^2 \right]^j }
\no
&=& i \, S_n \,
\frac{ (-p^2)^{n/2}}{(p^2)^{i+j}}
\, ( \Delta p )^m \, \frac{1}{\Gamma (i) \Gamma (j)}
\; \int_0^1 d x \; x^m \; x^{n/2 - 1 -i} (1-x)^{n/2-1-j}
\no
&&
\times \, \left[
\Gamma (i + j - \frac{n}{2}) x p^\mu +
\frac{1}{2} \, \lc \left(i+j-\frac{n}{2} \right) (1 - 2 \, x)
\right.
\right.
\no
&&
\left.
\left.
+ \Gamma (i+j-1-\frac{n}{2}) ( (1-x) (1-j) - x (1-i)) \rc
\frac{\Delta ^\mu p^2}{\Delta p} \right],
\label{B.2}
\\
I_{ij}^{\mu \nu} &=&
\int \frac{d^n q}{(2 \pi)^n} \,
\frac{q^\mu \, q^\nu \,
 (\Delta q)^m}{ \left[q^2\right]^i \left[(q-p)^2 \right]^j }
\no
&=& i \, S_n \,
\frac{ (-p^2)^{n/2}}{(p^2)^{i+j}}
\, ( \Delta p )^m \, \frac{1}{\Gamma (i) \Gamma (j)}
\, \int_0^1 d x \; x^m \; x^{n/2 - 1 -i} (1-x)^{n/2-1-j}
\no &&
\times \left[
\Gamma(i+j- \frac{n}{2} ) x^2 p^\mu p^\nu
+ \frac{1}{2} \, \Gamma \left( i + j - 1 - \frac{n}{2} \right)
x ( 1 - x) g^{\mu \nu} p^2 +
\right.
\no &&
+ \frac{\Delta^\mu p^\nu + p^\mu \Delta^\nu}{2 \, \Delta p} p^2
\lc
\Gamma \left(i+j-\frac{n}{2} \right) (x - 2 \, x^2)
\right.
\no
&&
+ \Gamma \left(i+j-1-\frac{n}{2} \right)
(-x (1-x) \, j + x^2 (i-1))
\no &&
+ \frac{\Delta^\mu \Delta ^\nu}{4 \, ( \Delta p )^2}
(p^2)^2 \lc
\Gamma \left(i+j-\frac{n}{2} \right) (1-4\, x + 4 \, x^2)
\right.
\no &&
+ 2 \, \Gamma (i + j -1 - \frac{n}{2})
 \left( (1-x)^2 (1-j) + x^2 (1-i)
+ x (1-x) (i+j-1)
\right)
\no &&
+ \Gamma \left(i+j-2-\frac{n}{2} \right)
\lc (1-x)^2 (j-1) (j-2) + x^2 (i-1) (i-2)
\right.
\no &&
\left.
\left.
- 2 \, x (1-x) (i-1) (j-1) \rc
\right],
\label{B.3}
\end{eqnarray}
where $S_n$ is the spherical factor $S_n = \pi^{\frac{n}{2}} / (2 \pi)^n$.

\noindent
The tensorial reduction of the two-loop tensor Feynman integrals
is much more complicated and has been performed by using the
program {\it FeynCalc} \cite{X31}.
The numerators of the two-loop Feynman integrals have the
following structure.
\begin{equation}
A_i = f_i^{\sigma_1 \cdots \sigma_k} \,
 \tilde{f}_{i \sigma_1 \cdots \sigma_k}(q_1, q_2),
\label{B.4}
\end{equation}
where $q_1$ and $q_2$ denote the integration momenta.
Explicit forms of $f_i^{\sigma_1 \cdots \sigma_k}$ and
$\tilde{f}_{i, \sigma_1 \cdots \sigma_k}(q_1, q_2)$ are
($\e^{\alpha \beta \sigma \delta} p_\alpha \Delta_\beta =
\e^{p \Delta \sigma \delta}$, etc.)
\begin{eqnarray}
f_1(q_1, q_2) &=& \e^{q_1 q_2 \Delta p} ,
\qquad \qquad \; \; \; \,
\tilde{f}_1(q_1,q_2) \; \; =\;  \; \e^{q_1 q_2 \Delta p} ,
\label{B.5}
\\
f_2^{\alpha}(q_1, q_2) &=& \e^{q_1 p \Delta \alpha} ,
\qquad \qquad \; \;
\tilde{f}_{2,\alpha} (q_1, q_2)\;   \; = \;\;  \e^{q_2 p \Delta \alpha} ,
\label{B.6}
\\
f_3^{\alpha \beta} ( q_1, q_2 ) &=& \e^{q_1 q_2 \alpha \beta} ,
\qquad \qquad
\tilde{f}_{3,\alpha \beta} (q_1, q_2)
\; \; = \;\;
{\rm Tr}( \qs_1 \qs_2 \Ds \ps \gamma_\alpha \gamma_\beta \gamma_5).
\label{B.7}
\end{eqnarray}
The tensor integrals can be represented as
\begin{eqnarray}
I^{ { \alpha; \alpha \beta ; \alpha \beta \sigma ;
      \alpha \beta \sigma  \delta
    }
  }(\Delta, p) &=&
\int \frac{d^n q_1}{(2 \pi )^n} \,
\int \frac{d^n q_2}{(2 \pi )^n} \,
K(q_1, q_2, \Delta,p)
\no
&&
\phantom{
\int \frac{d^n q_1}{(2 \pi )^n} \,
\int \frac{d^n q_2}{(2 \pi )^n} \,
}
\times
 \lc
q_i^\alpha; q_i^\alpha q_j^\beta;
q_i^\alpha q_j^\beta q_k^\sigma;
q_i^\alpha q_j^\beta q_k^\sigma q_l^\delta
\rc,
\label{B.8}
\end{eqnarray}
where $i=1,2$ and $j = 1,2$. Further we have the definition
\begin{eqnarray}
K(q_1, q_2, \Delta, p) &=&
\frac{(\Delta q_1)^a (\Delta q_2)^b (\Delta (p-q_1))^c (\Delta (p-q_2))^d
(\Delta (q_1 - q_2))^e}{(q_1^2)^f (q_2^2)^g ((q_1-p)^2)^h ((q_2 - p)^2)^i
((q_1-q_2)^2)^j} .
\label{B9}
\end{eqnarray}
Notice that the integers $a-g $ can take positive as well as negative integer
values.
By virtue of Lorentz covariance the integral $I(\Delta, p)$ can now be written
as
\begin{eqnarray}
I^{ { \alpha; \alpha \beta ; \alpha \beta \sigma ;
      \alpha \beta \sigma  \delta
    }
  }(\Delta, p) &=&
 \sum_s \lc
T_s^\alpha ,  T_s^{\alpha \beta}, T_s^{\alpha \beta \sigma},
 T_s^{\alpha \beta \sigma \delta}
\rc
\no
&& \times
\int \frac{d^n q_1}{(2 \pi )^n} \,
\int \frac{d^n q_2}{(2 \pi )^n} \,
\sum_r f_r(p^2,n) K_r(q_1, q_2, \Delta ,p) ,
\label{B.10}
\end{eqnarray}
with
\begin{eqnarray}
T_s^\alpha &=& \lc p^\alpha; \Delta^\alpha \rc,
\label{B.11} \\
T_s^{\alpha \beta} &=& \lc g^{\alpha \beta}; p^\alpha p^\beta; \Delta^\alpha
\Delta^\beta \rc,
\label{B.12} \\
T_s^{\alpha \beta \sigma} &=&
\lc g^{\alpha \beta} p^\sigma; \cdots; \Delta^\alpha \Delta^\beta \Delta^\sigma
\rc,
\label{B.13} \\
T_s^{\alpha \beta \sigma \delta} &=&
\lc g^{\alpha \beta} g^{\sigma \delta}; \cdots; \Delta^\alpha \Delta^\beta
\Delta^\sigma \Delta^\delta \rc,
\label{B.14}
\end{eqnarray}
where the $K_r$ are of the same type as the $K$ in (\ref{B.8}) (but with
different
indices a-j) and the $f_r(p^2,n)$ are simple polynomial-like
functions determined by the tensorial reduction.

\noindent
In this way all Lorentz indices are transformed away from the integration
momenta to the external
momentum $p$ and the lightlike vector $\Delta$. The advantage of the tensorial
reduction method is
revealed when one evaluates e.g.~the expression $\tilde{f}_{3,\alpha \beta}$
(\ref{B.7}). This gets
simplified to
Tr$(\Ds \ps \gamma_\alpha \gamma_\beta \gamma_5) =
- 4 \, i \, \e^{\Delta p \alpha \beta}$.
Hence one can avoid
any $\gamma_5$-prescription dependence arising
from the non-unique way of calculating a trace of
six $\gamma$-matrices plus the
$\gamma_5$-matrix
in $n$ dimensions.
The explicit reduction formulae, which are too lengthy to be presented here,
are obtained by using projection methods. They are incorporated in the program
{\it FeynCalc 3.0} \cite{X25}.
The scalar integrals which appear on the right hand side of (\ref{B.10}) are
calculated in
\cite{X10} using the algebraic manipulation program FORM \cite{X27}. The
two-loop integrals including
the tensorial reduction have been checked by recalculating the spin averaged
splitting functions which
have been computed in the past (see \cite{X14}-\cite{X17})
and we found full agreement.
\appendixoff

\vspace*{1.0cm}
\noindent

\clearpage
\afterpage{\clearpage}

{\bf Figure captions.}\\
\begin{description}
\item[Fig. 1] One-loop graphs contributing to the spin OMEs;
(a), (b):
$A_{NS,qq}^{(1)}$, $A_{S,qq}^{(1)}$;
(c): $A_{S,gq}^{(1)}$; (d), (e):
$A_{S,qg}^{(1)}$; (f), (g): $A_{S,gg}^{(1)}$.
Graphs with external self-energies and with triangular
fermion-loops
where the arrows are reversed have been included in the
calculation but are not  shown in the figure.
Graphs which are not symmetric with respect to the vertical line
through the operator vertex have to be counted twice.
\item[Fig. 2]
Two-loop graphs contributing to the spin non-singlet OME
$A_{NS,qq}^{(2)}$. Graphs with external self-energies
have been included in the calculation but are not drawn in
the figure.
Graphs which are not symmetric with respect to the vertical line
through the operator vertex have to be counted twice.
\item[Fig. 3]
Two-loop graphs contributing to the spin pure-singlet
OME $A_{PS,qq}^{(2)}$. Graphs with triangular fermion loops where the arrows
are
reversed have been included  in the calculation but are not shown in
the figure.
Graphs which are not symmetric with respect to the vertical line
through the operator vertex have to be counted twice.
\item[Fig. 4]
Two-loop graphs contributing to the spin singlet OME $A_{S,qg}^{(2)}$.
Graphs with triangular fermion loops where the arrows are reversed and
diagrams containing external self energies have been
included but are not shown in the figure.
Graphs which are not symmetric with respect to the vertical line
through the operator vertex have to be counted twice.
\item[Fig. 5]
Two-loop graphs contributing to the spin singlet OME $A_{S,gq}^{(2)}$.
Graphs with external self-energies
have been included in the calculation but are not drawn in
the figure.
Graphs which are not symmetric with respect to the vertical line
through the operator vertex have to be counted twice.
\item[Fig. 6]
Two-loop graphs contributing to the spin singlet OME
$A_{S,gg}^{(2)}$. Graphs with external
self-energies and diagrams with ghost and triangular fermion loops
where the arrows are reversed have been included in the
calculation but are not drawn in the figure.
Graphs which are not symmetric with respect to the vertical line
through the operator vertex have to be counted twice.
\end{description}

\clearpage

\begin{figure}[p]
\centering
\includegraphics[width=0.7\textwidth]{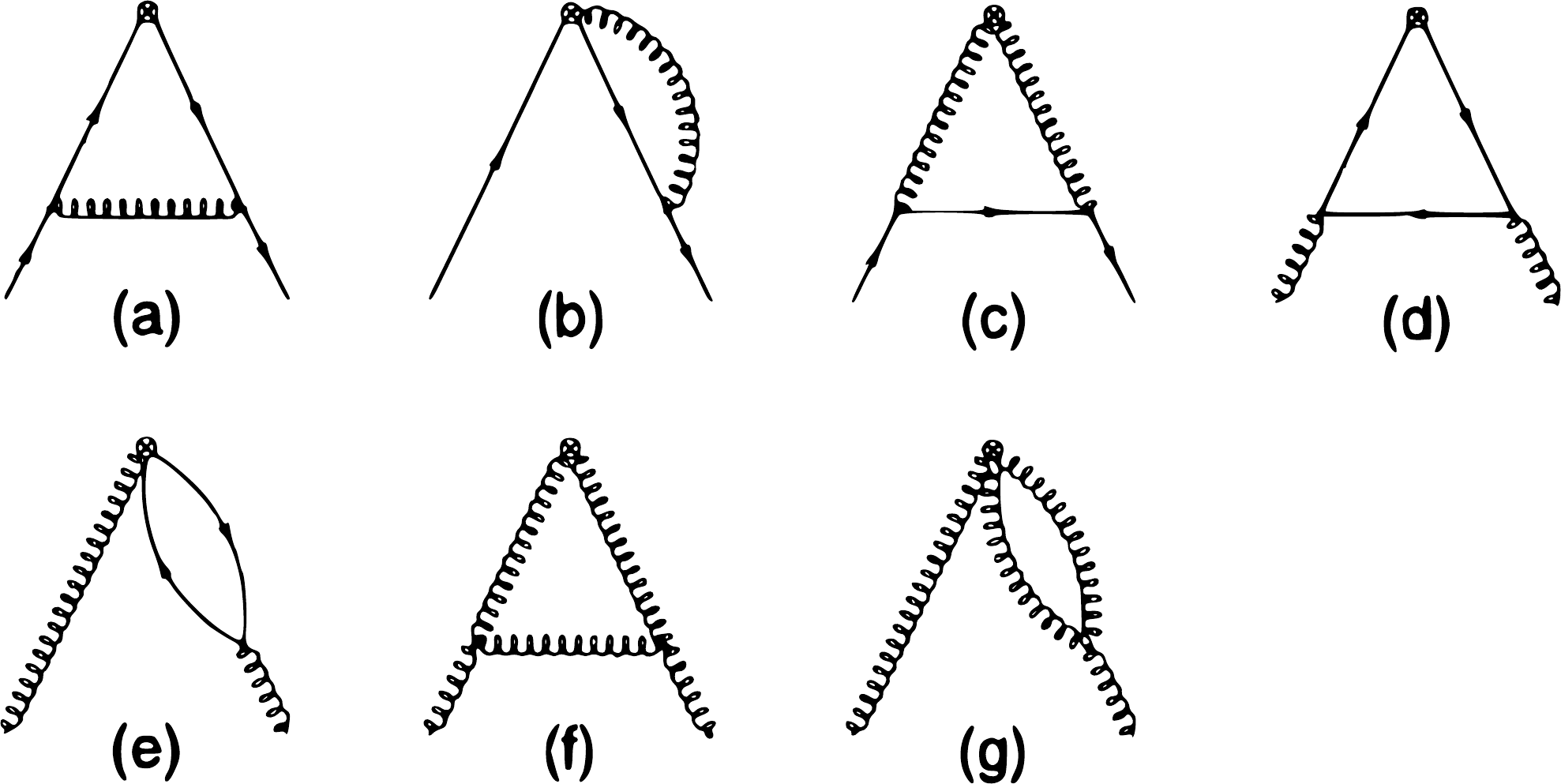}
\caption{}
\label{fig1}
\end{figure}
\clearpage
\afterpage{\clearpage}

\begin{figure}[p]
\centering
\includegraphics[width=0.9\textwidth]{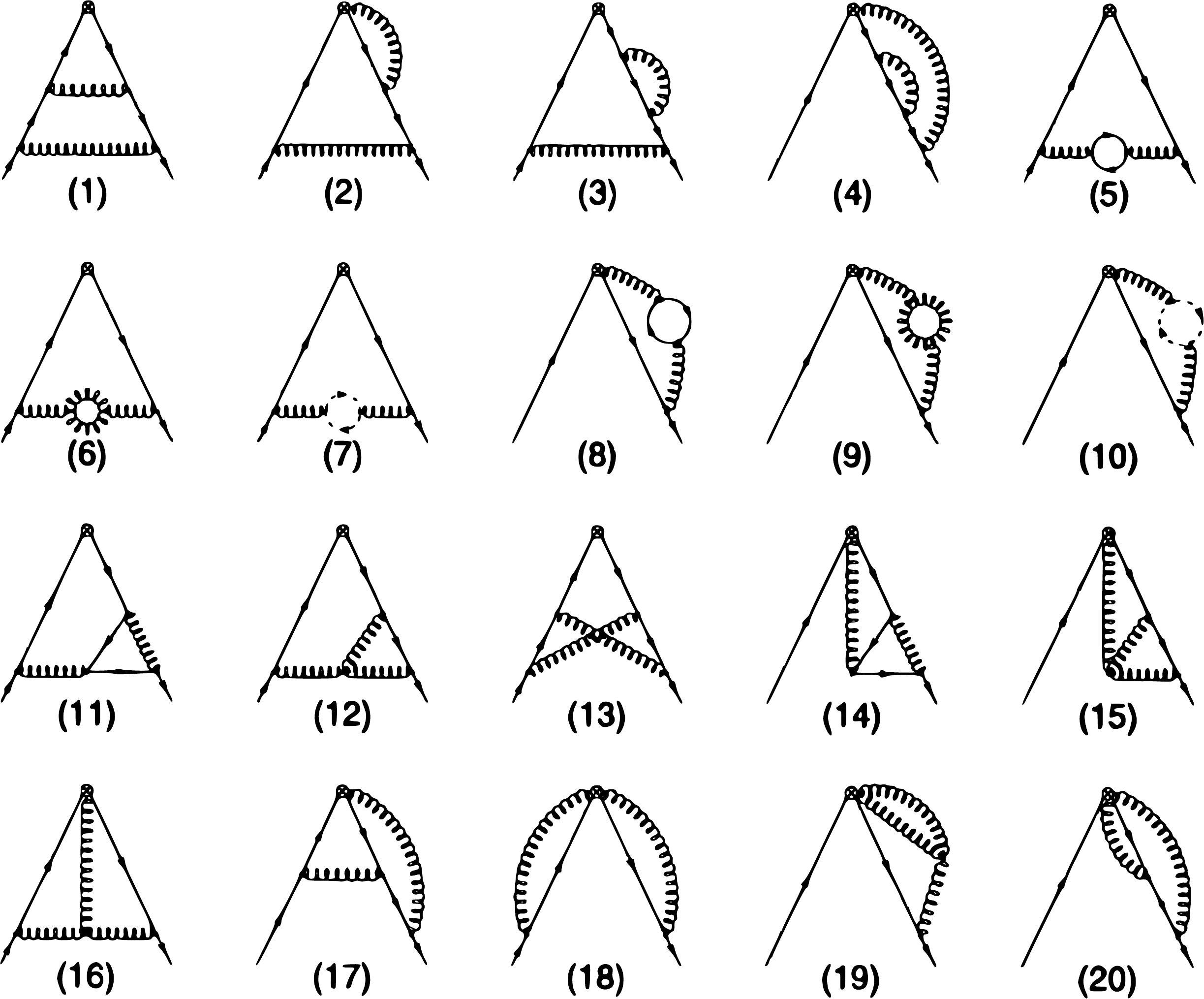}
\caption{}
\label{fig2}
\end{figure}

\clearpage
\afterpage{\clearpage}

\begin{figure}[p]
\centering
\includegraphics[width=0.6\textwidth]{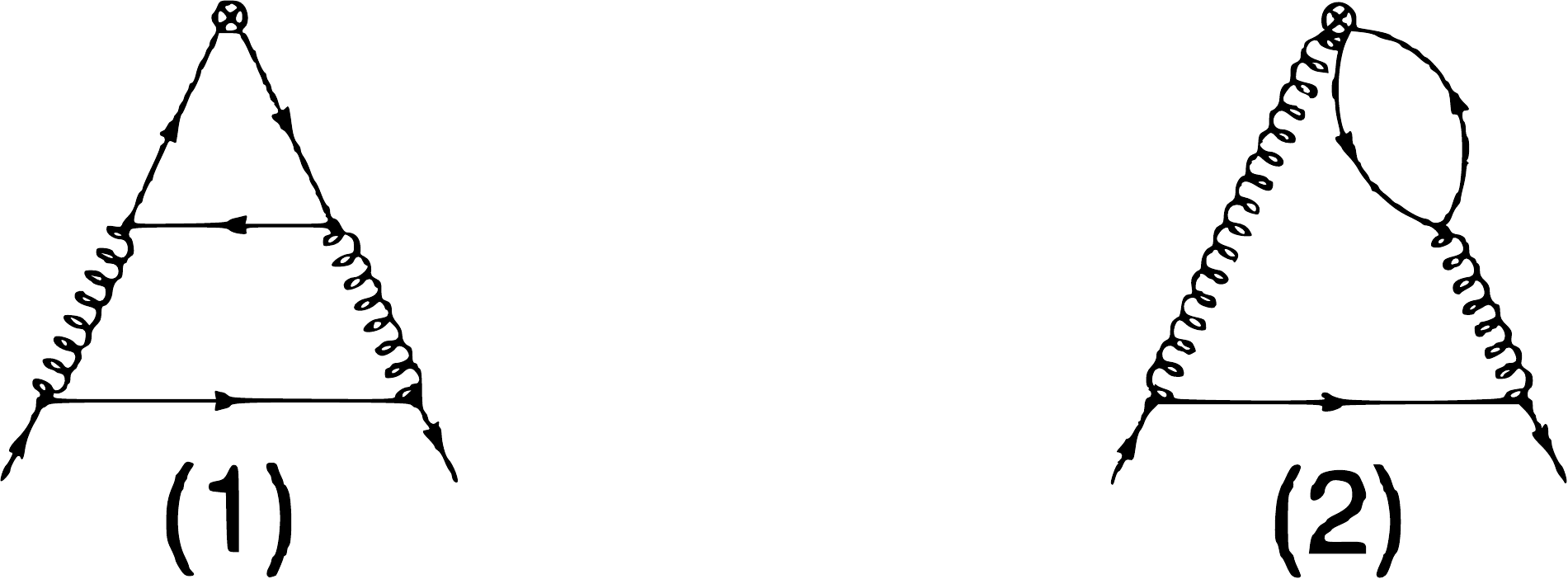}
\caption{}
\label{fig3}
\end{figure}
\clearpage
\afterpage{\clearpage}

\begin{figure}[p]
\centering
\includegraphics[width=0.9\textwidth]{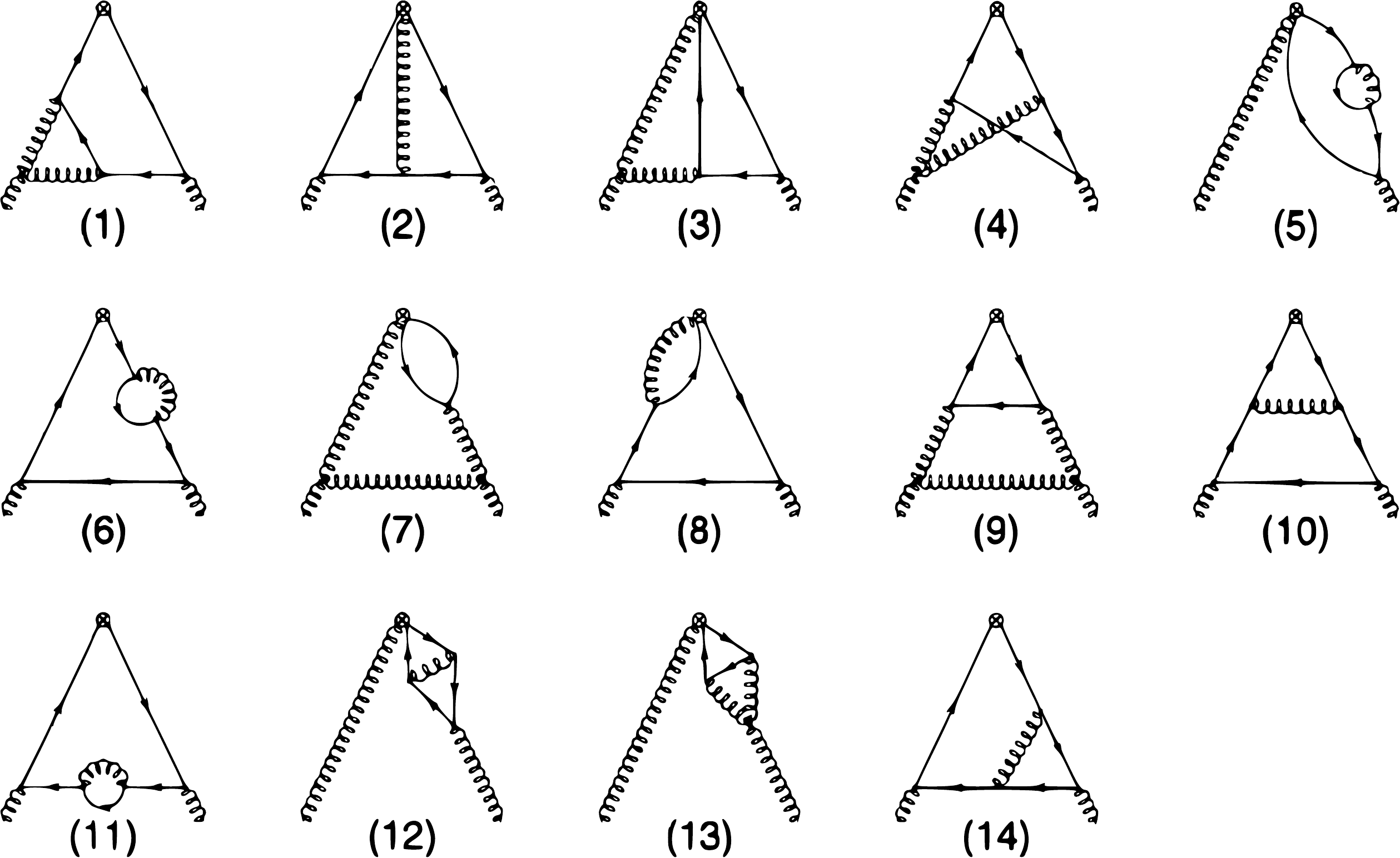}
\caption{}
\label{fig4}
\end{figure}
\clearpage
\afterpage{\clearpage}

\begin{figure}[p]
\centering
\includegraphics[width=0.9\textwidth]{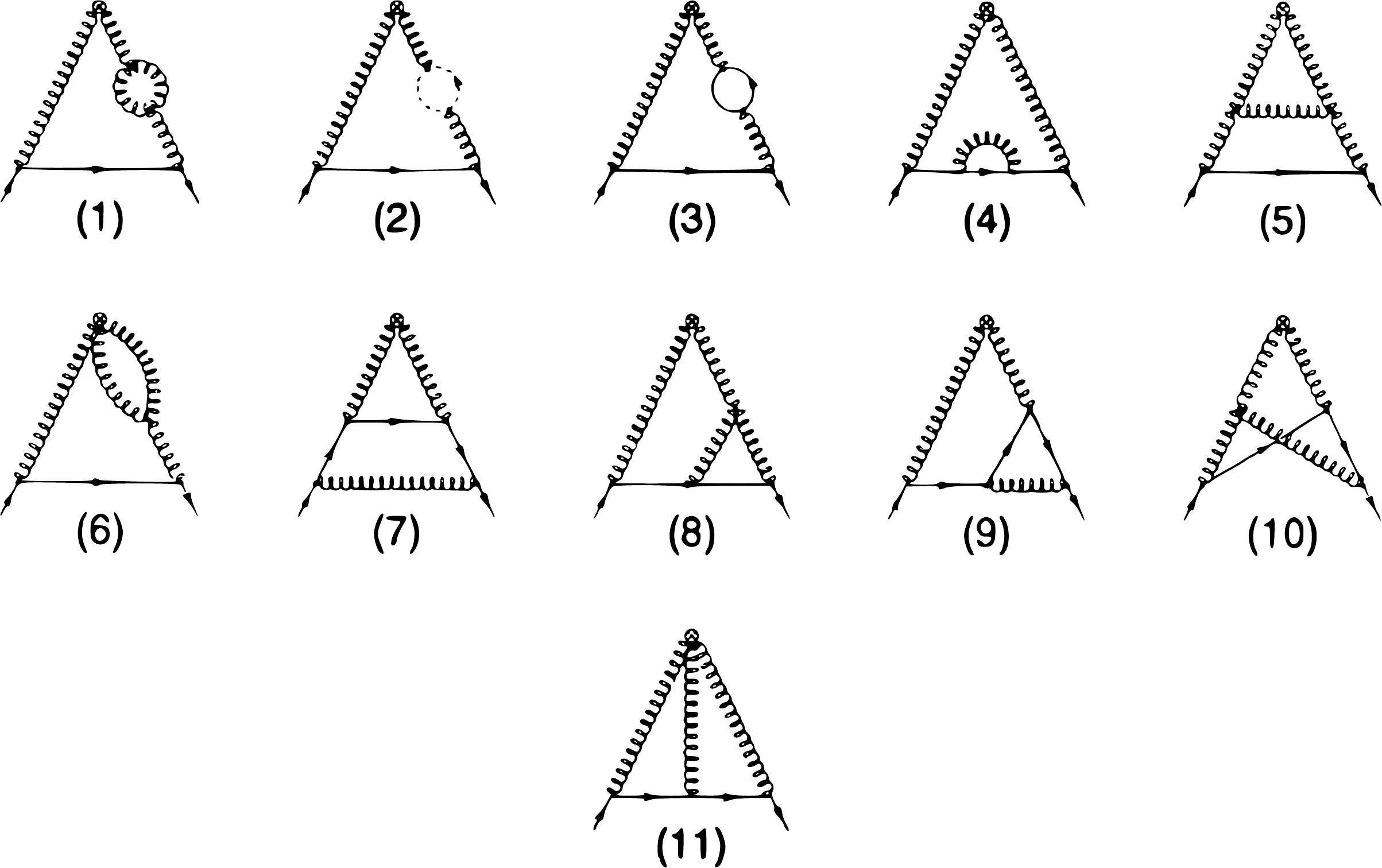}
\caption{}
\label{fig5}
\end{figure}
\clearpage
\afterpage{\clearpage}

\begin{figure}[p]
\centering
\includegraphics[width=0.9\textwidth]{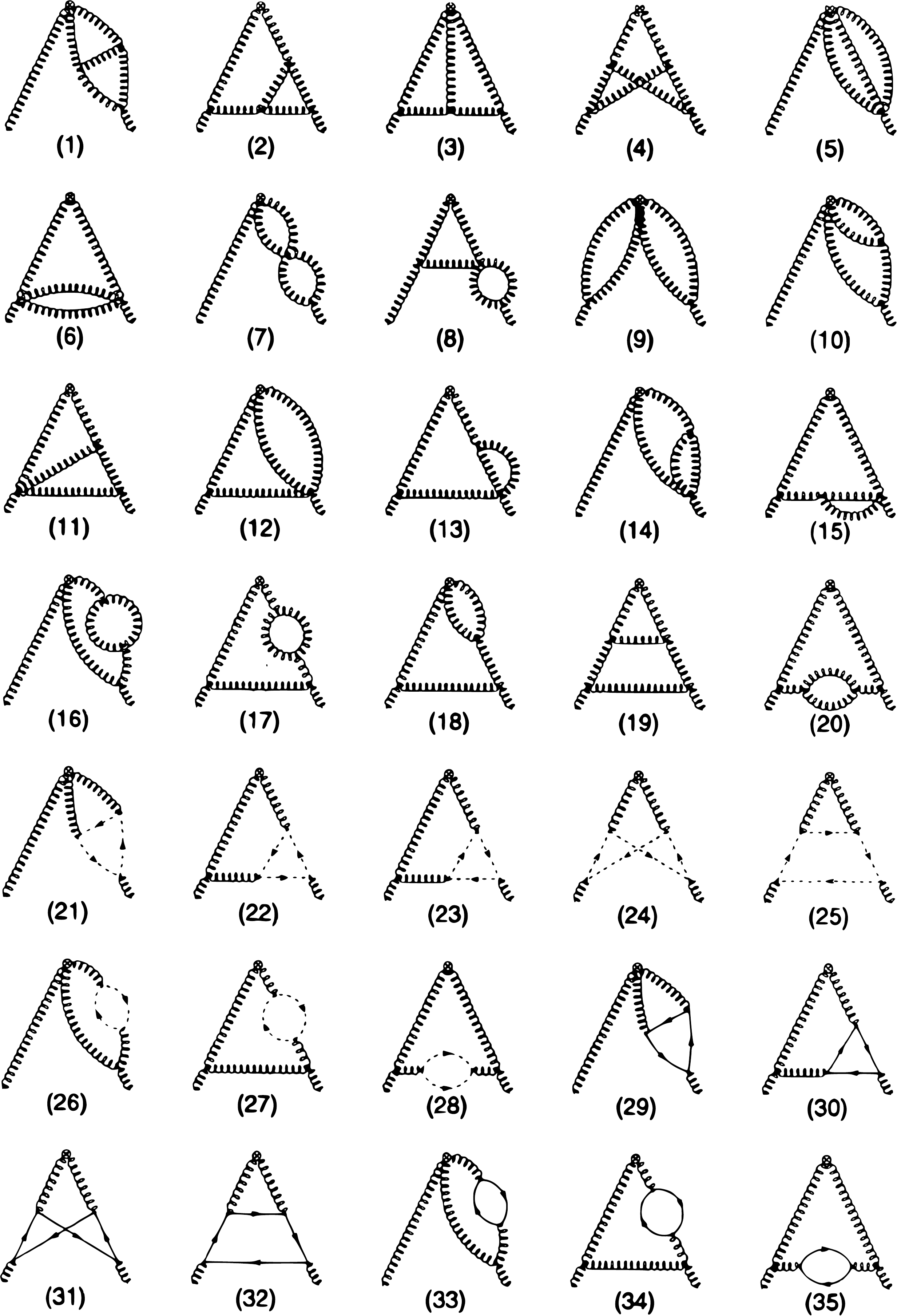}
\caption{}
\label{fig6}
\end{figure}

\end{document}